\documentclass[
reprint,
nofootinbib,
amsmath, 
amssymb,
aps,]{revtex4-2} 
\usepackage{subfig, caption}
\usepackage{natbib}
\usepackage{mathtools}
\usepackage{txfonts}
\usepackage{scalerel}
\usepackage{graphicx}  
\usepackage{bm}        
\usepackage{amssymb}   
\usepackage{lipsum}
\usepackage{lineno}
\usepackage{amsmath,array}
\usepackage{caption}
\usepackage{hyperref}
\hypersetup{
	colorlinks=true,
	linkcolor=blue,
	filecolor=magneta,      
	urlcolor=blue,
}

\DeclareCaptionJustification{justified}{\leftskip=0pt \rightskip=0pt \parfillskip=0pt plus 1fil}

\usepackage{tikz}
\usetikzlibrary{svg.path}
\definecolor{kjm}{HTML}{A6CE39}
\tikzset{
    kz/.pic={
        \fill[kjm] svg{M256,128c0,70.7-57.3,128-128,128C57.3,256,0,198.7,0,128C0,57.3,57.3,0,128,0C198.7,0,256,57.3,256,128z};
        \fill[white] svg{M86.3,186.2H70.9V79.1h15.4v48.4V186.2z}
        svg{M108.9,79.1h41.6c39.6,0,57,28.3,57,53.6c0,27.5-21.5,53.6-56.8,53.6h-41.8V79.1z M124.3,172.4h24.5c34.9,0,42.9-26.5,42.9-39.7c0-21.5-13.7-39.7-43.7-39.7h-23.7V172.4z}
        svg{M88.7,56.8c0,5.5-4.5,10.1-10.1,10.1c-5.6,0-10.1-4.6-10.1-10.1c0-5.6,4.5-10.1,10.1-10.1C84.2,46.7,88.7,51.3,88.7,56.8z};
    }
}
\newcommand\axc[1]{\href{https://orcid.org/#1}{\mbox{\scalerel*{
                \begin{tikzpicture}[yscale=-1,transform shape]
                \pic{kz};
                \end{tikzpicture}
            }{|}}}}

\begin{document}
\preprint{APS/123-QED}
\title{Constraining self-interacting ultrahigh-energy { muon} neutrinos by cosmic microwave background spectral distortion}
\author{Pravin Kumar Natwariya\,$^{\axc{0000-0001-9072-8430}}$\,}
\email{pvn.sps@gmail.com}
\affiliation{School of Fundamental Physics and Mathematical Sciences, Hangzhou
Institute for Advanced Study, University of Chinese Academy of Sciences (HIAS-UCAS), Hangzhou, 310024, China}
\affiliation{University of Chinese Academy of Sciences, Beijing, 100190, China}
\affiliation{International Centre for Theoretical Physics Asia-Pacific (ICTP-AP), Beijing, 100190, China}
\author{Shibsankar Si$^{\axc{0009-0001-8038-976X}}$\,}
\email{p21ph001@nitm.ac.in}
\affiliation{National Institute of Technology, Meghalaya, Shillong, Meghalaya 793003, India}
\author{Alekha C. Nayak$^{\axc{0000-0001-6087-2490}}$\,}%
\email{alekhanayak@nitm.ac.in}
\affiliation{National Institute of Technology, Meghalaya, Shillong, Meghalaya 793003, India}%
\author{Tripurari Srivastava$^{\axc{0000-0001-6856-9517}}$\,}%
\email{tripurarisri022@gmail.com}
\affiliation{Institute of Particle Physics and Key Laboratory of Quark and Lepton Physics (MOE), Central China Normal University, Wuhan, Hubei 430079, China}%
\affiliation{Department of Physics and Astrophysics, University of Delhi, Delhi, 110007, India}
\date{\today}
\begin{abstract}
\centerline{\bf Abstract}
The neutrino telescopes have firmly established the existence of ultrahigh-energy neutrinos. Observations of these neutrinos offer a unique probe of neutrino self-interactions. This work investigates how the self-interacting neutrinos, mediated by scalar bosons, inject energy into the medium through radiative scattering with the cosmic neutrino background, leaving an imprint on the cosmic microwave background (CMB) spectrum. The energy injection into plasma in redshift ranges, $5\times10^4\lesssim z\lesssim2\times10^6$ and $ z\lesssim5\times10^4$, leads to $\mu$-type and \textit{y}-type CMB spectral distortions, respectively. Using observational constraints from Cosmic Background Explorer/Far Infrared Absolute Spectrophotometer (COBE/FIRAS) and projected sensitivities from Primordial Inflation Explorer (PIXIE) experiments for $\mu$-type and $y$-type CMB distortions, we derive the stringent upper bounds on the self-interaction coupling strength as a function of mediator mass for neutrino interactions. We focus on flavor-specific self-interaction related to { muon} neutrinos and sub-GeV mass mediators ($m_{\phi}$). We find the upper bound on the self-interaction coupling strength to be {$\sim 2.8\times 10^{-4}$} for the muon neutrino, considering ultrahigh-energy muon neutrino energy to be 1~PeV and PIXIE projected upper bounds on $y$-type CMB spectral distortion. The bound remains constant till the mediator mass reaches the center-of-mass energy, and after that, it gets relaxed and becomes proportional to the mediator mass. We have also compared our results with existing bounds in the literature. Our findings indicate that CMB spectral distortion could play a decisive role in exploring neutrino physics beyond the standard model of particle physics, and future missions like PIXIE can provide valuable insights.

\end{abstract}
\maketitle
\section{Introduction}

The advancement in understanding the physical properties and interactions of neutrinos has been a cornerstone of particle physics and modern cosmology. Neutrinos are among the most abundant particles in the Universe and provide a potential window to explore physics beyond the Standard Model. They can also play a significant role in multimessenger astronomy to study exotic astrophysical phenomena, as well as in understanding the evolution of the Universe. In multimessenger astronomy, the neutrinos are particularly interesting because they travel in straight lines from their sources, as they are nearly massless particles and rarely interact with matter, passing through dense environments that block other types of probes, such as cosmic rays. This makes them ideal for probing extreme cosmic phenomena, such as supernovae, gamma-ray bursts (GRBs), blazars, and neutron star mergers \cite{Mszros:2019, Salesa:2021, Guepin:2022, Barenboim:2019tux, Sharma:2024}.

In the Standard Model of particle physics, neutrinos are massless and interact only through the weak force. However, the discovery of neutrino oscillations has firmly established that neutrinos possess nonzero masses, compelling physicists to explore beyond the Standard Model of particle physics  \cite{Pontecorvo:1958, Maki:1962, Pontecorvo:1968, Bahcall:1976, Smirnov:2006, Lesgourgues:2006, Super-K:2004, Super-K:2005, SNO:2005, SNO::2005}. Neutrinos have been observed across a wide range of energies, including ultrahigh-energy (UHE) neutrinos with energies in the PeV range, first detected by the IceCube Neutrino Observatory and more recently by the KM3NeT (Kilometre cube Neutrino Telescope) Collaboration \cite{IceCube:2020wum, IceCube:2023ame, IceCube:2023lxq, Aiello2025}. The IceCube Collaboration reported observation of the flux of high-energy neutrinos from the Galactic plane by analyzing 10 years of neutrino emission data using machine learning techniques \cite{IceCube:2023ame, IceCube:2023lxq}. These observations have firmly established the existence of high-energy neutrinos. The detection of such energetic cosmic neutrinos opens a unique window into the extreme Universe and offers a potential probe for nonstandard interactions. 

The IceCube High-Energy Starting Event (HESE) 7.5-year data had nontrivial spectral features \cite{IceCube:2020wum, Esteban:2021tub}. The astrophysical acceleration mechanism spectra are expected to follow a power law as a function of energy. The deviation in the HESE observed spectrum from the power law can be explained by introducing strongly self-interacting neutrinos \cite{Esteban:2021tub, Carpio:2021jhu, Bustamante:2020mep, Mazumdar:2020ibx}. Moreover, strong neutrino self-interaction ($\nu$SI) mediated by scalar or vector bosons has gained a lot of attention as it can allow KeV-scale sterile neutrinos to be a viable dark matter candidate in the context of the famous Dodelson-Widrow mechanism \cite{DeGouvea:2019wpf, Kelly:2020pcy, Benso:2021hhh, Dhuria:2023yrw}. 
Additionally, introducing moderately strong self-interactions offers a framework to address several long-standing anomalies, including the Hubble tension \cite{Kreisch:2019yzn, RoyChoudhury:2020dmd}, and discrepancy in the anomalous magnetic moment of the muon, \((g-2)_\mu\) \cite{PhysRevD.105.L051702}. In Ref. \cite{PhysRevD.109.063007}, the author claims that recent gamma-ray events GRB221009A can be explained by the interaction of high-energy neutrinos with CMB neutrinos through nonstandard self-interaction mediated by light scalar bosons. 

There is a possibility that these UHE neutrinos may also originate from decaying or annihilating superheavy dark matter \cite{Guepin:2021qai, Berghaus:2025jwb}. In Ref. \cite{Kachelriess:2018rty}, the authors consider IceCube datasets \cite{IceCube:2016uab}, especially two events with 2.6 PeV and 2.7 PeV energy, and claim to be the origins of superheavy dark matter. The authors of the article \cite{Bhattacharya:2017jaw} analyze 4-year IceCube HESE data and argue that a significant contribution from dark matter decay is always slightly favored to explain the spectrum. The authors also explore the scenario where all the data are interpreted by dark matter decays only and conclude that this scenario better fits HESE data than the isotropic power-law flux. The IceCube Collaboration also explored decaying superheavy dark matter scenarios with six years of IceCube data, focusing on muon neutrino ``track" events from the Northern Hemisphere, and showed that the observed high-energy neutrino flux can be explained by a combination of a dark matter component and a diffuse astrophysical flux \cite{IceCube:2018tkk}. In the article \cite{Sui:2018bbh}, the authors also favored the astrophysical plus dark matter interpretation over the purely astrophysical explanation of the IceCube 6-year HESE data and 8-year throughgoing muon events above 10 TeV. The next-generation UHE neutrino telescopes, such as IceCube-Gen2, by exploring the angular and energy distributions of detected events \cite{IceCube-Gen2:2020qha, Fiorillo:2023clw}, GRAND (Giant Radio Array for Neutrino Detection) \cite{GRAND:2018iaj}, POEMMA (Probe of Extreme Multi-Messenger Astrophysics) \cite{POEMMA:2020ykm}, RNO-G (Radio Neutrino Observatory in Greenland) \cite{RNO-G:2020rmc}, and others \cite{Otte:2018uxj, Neronov:2019htv}, will be able to explore the possible connection between UHE neutrinos and superheavy dark matter. In this article, we consider ultrahigh-energy neutrinos from dark matter and constrain the parameter space of neutrino self-interaction mediated by scalar bosons using the CMB spectral distortion.

The parameter space of neutrino self-interaction has been studied in various earlier articles; in the article \cite{Ioka:2014kca}, the authors explore the PeV-scale neutrinos in the context of interactions of these high-energy neutrinos with the cosmic neutrino background (C$\nu$B), and set an upper bound on the coupling to be $ g < 3\times10^{-2}$ for the mediating boson mass to be $\lesssim2$~MeV. In the article \cite{Oldengott:2017fhy}, the authors analyze the effects of the interactions on the cosmic microwave background anisotropies and find the upper bound to be $G_{\rm eff}\lesssim 10^{-3}~{\rm MeV^{-2}}$, where $G_{\rm eff}\equiv g^2/m_\phi^2$, and $m_\phi$ is the mass of the mediator. The self-interaction coupling parameter, $G_{\rm eff}= 10^{-3.9^{+1.00}_{-0.93}}~{\rm MeV^{-2}}$, may imply a higher value of the Hubble parameter, reducing the discrepancy between the CMB measurement and local measurements (low-redshift observations) \cite{Kreisch:2019yzn}. Such large values of $G_{\rm eff}$ for flavor-universal interactions are excluded by the big-bang nucleosynthesis
(BBN)-only bounds on $\Delta N_{\rm eff}$ and by laboratory searches for rare kaon ($K$) decays and neutrinoless double-beta decay \cite{Blinov:2019gcj}. However, the flavor-dependent scenario may viably reduce the Hubble tension \cite{Blinov:2019gcj}. The bounds using  SN1987A data on the parameter space for light bosons coupled to neutrinos have been explored in the articles \cite{Fiorillo:2022cdq, Chang:2022aas, Fiorillo:2023ytr, Fiorillo:2023cas, Telalovic:2024cot}. In the article \cite{PhysRevD.110.123033}, the authors obtain the upper bounds of the order of  $\mathcal{O}(10^{-4})$ on the coupling parameter $g$ by analyzing the impact of self-interacting neutrinos on the 21-cm signal during the cosmic dawn. The authors consider the production of high-energy neutrinos from the decay of superheavy dark matter. These high-energy neutrinos can interact with the cosmic neutrino background (C$\nu$B), subsequently producing energetic photons and affecting the {temperature} evolution. The coupling parameter $g > \mathcal{O}(10^{-2})$, for sub-MeV mediator mass, is excluded for self-interacting $\tau$ neutrinos by currently available IceCube HESE 7.5 years of data \cite{Esteban:2021tub}. The IceCube-Gen2 will be able to detect the coupling strength up to $\mathcal{O}(10^{-4})$ for the MeV mass range mediator \cite{Esteban:2021tub}.

The CMB spectral distortion offers a unique and robust probe of the thermal history of the early Universe and the presence of any new physics. The CMB spectrum is well-approximated by the blackbody spectrum \cite{Fixsen:1996nj, Mather:1993ij}. While the adiabatic expansion of the Universe preserves the CMB blackbody spectrum by uniformly redshifting photon energies, deviations arise from nonthermal processes such as exotic energy injections or dissipation processes. These distortions are mainly categorized as $\mu$-type or $y$-type, distinguished by the redshift of energy injection \cite{khatri2012beyond, Kunze_2014}. Energy injection into the plasma in the redshift range of $5\times10^4\lesssim z\lesssim2\times10^6$ results in $\mu$-type of CMB spectral distortions, while the energy injection below a redshift of $5\times10^4$ results in $y$-type of CMB spectral distortions. The COBE/FIRAS (Cosmic Background Explorer / Far Infrared Absolute Spectrophotometer) set upper bounds on the \textit{y}-type ($\mu$-type) distortion to be $y = 1.5\times10^{-5}$ $(\mu = 9.0 \times10^{-5})$ with 95\% confidence level \cite{Fixsen:1996nj}. The future experiment Primordial Inflation Explorer (PIXIE) aims to detect $y$-type ($\mu$-type) distortion at levels of $y = 10^{-8}$ ($\mu=5\times10^{-8}$) at a $5\sigma$ level \cite{Kogut:2011xw}. In the present paper, we use these limits on CMB spectral distortions to derive the upper bound on the neutrino self-interaction coupling strength in the presence of radiative scattering of ultrahigh-energy neutrinos with the cosmic neutrino background mediated by scalar bosons. 

The structure of the paper is as follows: In section \ref{self-interaction}, we present a brief overview of the self-interaction of the UHE neutrinos mediated by scalar bosons.  In section \ref{energy-injection}, we explain how the interaction of UHE neutrinos and C$\nu$B injects the energy into the medium. How this energy injection into the plasma affects the $\mu$ and $y$ types CMB spectral distortion is explained in section \ref{cmbsd}. We discuss results in section \ref{Result}. Finally, we conclude the results in section \ref{Conclusion}.

\begin{figure*}[ht]
	\begin{center}
		\includegraphics[width=0.8\textwidth]{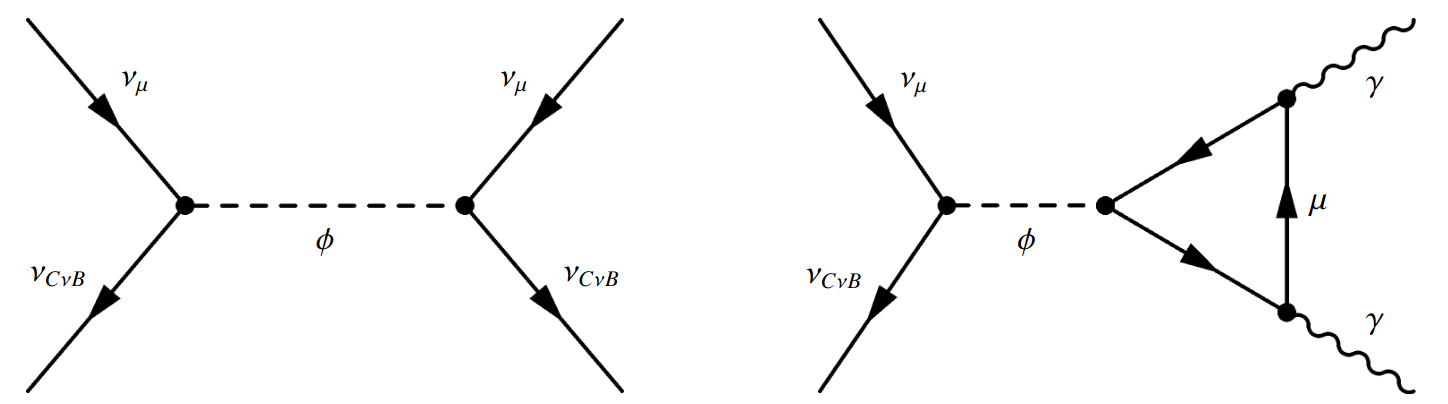}
	\end{center}
	\caption{Feynman diagrams (a) for scattering of ultrahigh energy neutrinos ($\nu_\mu$) with cosmic neutrino background ($\nu_{C\nu B}$) mediated by a scalar $\phi$ and (b) radiative scattering of ultrahigh energy neutrinos with cosmic neutrinos background into ultra high energy photons mediated by scalar propagator and fermion (muon) in a loop.}
	\label{fig:feynman}
\end{figure*}

\section{ultrahigh-Energy Neutrinos and Self-interaction } \label{self-interaction}

As we mentioned in the introduction, the origin of UHE neutrinos from decaying superheavy dark matter cannot be completely ruled out \cite{Guepin:2021qai, Berghaus:2025jwb, Kachelriess:2018rty, Bhattacharya:2017jaw, IceCube:2018tkk, Sui:2018bbh}. We examine a theoretical framework in which a superheavy dark matter particle, with mass $m_\chi\geq \mathrm{PeV}$, predominantly decays into ultrahigh energy neutrinos. The resulting neutrino yield is determined by the specific decay channels of the dark matter particle. In scenarios where dark matter mainly undergoes a two-body decay into a UHE neutrino-antineutrino pair $\chi\rightarrow\nu\Bar{\nu}$, the resulting neutrino flux will be characterized by an energy $E_{\nu}\approx m_\chi/2$. 
These UHE neutrinos can have flavor-specific strong interactions with background cosmic neutrinos through a new hidden sector mediated by a scalar or vector boson~\cite{PhysRevD.109.063007}. Such interactions can maintain thermal equilibrium among neutrinos for longer periods. This interaction has a notable impact on the CMB spectrum and has been extensively studied in the context of the Hubble tension~\cite{PhysRevLett.123.191102, Oldengott:2017fhy, Kreisch:2019yzn}. In this work, we focus on self-interaction between muon-flavored neutrinos. Our choice to focus solely on the muon sector is partly inspired by potential new physics hinted at in muon-related observables such as the muon $g-2$ anomaly~\cite{Lindner:2016bgg, Capdevilla:2021kcf} and contribution to rare meson decays, such as $K \rightarrow \mu \nu \phi$~\cite{PhysRevLett.124.041802, Blinov:2024gcw}, which offer complementary experimental probes~\cite{Caputo:2021rux, Forbes:2022bvo, Jho:2019cxq}.
We consider a framework of self-interacting neutrinos mediated by a scalar that couples to the muon and muon neutrinos. Below the electroweak scale, these interactions can typically be described through the following Lagrangian \cite{PhysRevD.110.123033}:
\begin{eqnarray}    
    {\cal L } \supset  g_{\nu_\mu}{\bar\nu_\mu}{\nu_\mu}\,\phi + g_{\mu }\,{\bar \mu}{\mu}\,\phi.
    \label{eq:lag}
\end{eqnarray}
In this work, we treat neutrinos as Majorana particles. Weyl notation is used to describe the coupling of ultrahigh-energy muon neutrinos to the scalar mediator, while Dirac notation is employed for the scalar-muon interaction. The coupling constant $g_{\nu_\mu}$ is generally linked to the mechanism responsible for neutrino mass generation, such as seesaw models or spontaneous lepton number violation. In contrast, $g_\mu$ typically arises from nonrenormalizable operators involving both the Higgs and scalar fields~\cite{Zhu:2021vlz, Cesarotti:2023udo}, and may emerge naturally within ultraviolet (UV) completions that include extended scalar or fermion sectors. Below the electroweak scale, these interactions can be effectively described by Yukawa-type couplings of the scalar to muons and muon neutrinos. 
{While the precise relationship between \( g_{\nu_\mu} \) and \( g_\mu \) is model dependent, we consider two simplified benchmark choices for their ratio to illustrate how the resulting constraints depend on the relative strength of these couplings.}

The Lagrangian introduces two Yukawa-type interactions involving a new scalar mediator $\phi$: one allowing self-scattering of muon neutrinos, and the other coupling the scalar to muons. The neutrino self-interaction term $g_{\nu_\mu} \bar{\nu}_\mu \nu_\mu \phi$ is allowed due to the Majorana nature of the neutrinos and can lead to significant modifications in early-Universe neutrino behavior. The muon coupling term $g_\mu \bar{\mu} \mu \phi$, allows the scalar to interact with charged leptons of muon flavor. While we do not explore the ultraviolet origin of these couplings, this setup provides a minimal and phenomenologically motivated scenario. 
However, we do not pursue that direction further, and instead concentrate on constraining this minimal coupling structure using astrophysical and cosmological data.

In this paper, we explore phenomenological consequences of the strong interaction of cosmic neutrinos with astrophysical neutrinos. 
In this framework, the cosmic background neutrinos can undergo scattering with astrophysical neutrinos via s-channel scattering processes having the same flavor as shown in the left diagram in Fig. \ref{fig:feynman}. This process can lead to photon pair production through a one-loop diagram involving a charged fermion, such as the muon ($\mu$), as shown in the right diagram in Fig. \ref{fig:feynman}. 
The emitted photons via radiative scattering can heat the plasma, which can significantly impact the CMB spectrum. 
The interaction cross-section of the UHE neutrinos with the cosmic neutrino background is given by \cite{PhysRevD.100.123007, PhysRevD.93.023509, PhysRevD.109.063007, PhysRevD.110.123033}
\begin{eqnarray}\label{eq:crossec}
     \sigma= \frac{81 \alpha^2 s}{4 \pi^3} \frac{ (g_{\mu }g_{\nu_\mu })^2 }{ m^2_\phi \Gamma^2_\phi  + (s - m^2_\phi)^2 }
     \times \left|1+ Q^2_\mu m^2_{\mu} C_0 \right|^2,
\end{eqnarray}
where, $\alpha\approx1/137$ is the fine-structure constant. $s$ is the center-of-mass energy, and it is defined as $s\approx{2 E_{\nu_\mu} E_{\nu}}$ \cite{Blinov:2019gcj}. $m_{\nu_\mu}\approx 0.1~\rm{eV}$ is the mass of the background neutrinos \cite{10.1093/mnras/stu1780}. $m_\phi$ is the mass of the mediator, $\Gamma_\phi$ is the decay width of the mediator, and is defined as $\Gamma_\phi= m_\phi g_{\mu }g_{\nu_{\mu }}/4\pi$. $m_\mu$ and $Q_\mu$ are the mass and charge of the muon, respectively. $C_0$ is the scalar Passarino-Veltman function given by \cite{PhysRevD.93.023509},
\begin{eqnarray} \label{eq:crossec1}
    C_0 (s, m_\mu) = \frac{1}{2 s} {\rm ln^2}\left|\frac{\left({1-  \frac{4 m^2_\mu}{s}}\right)^{1/2}-1}{\left({1-  \frac{4 m^2_\mu}{s}}\right)^{1/2}+1} \right|.
\end{eqnarray}

\section{Energy Injection into plasma due to Self-interaction of Ultrahigh-Energy Neutrinos} \label{energy-injection}

{
In the previous section (Section \ref{self-interaction}), we discussed how UHE neutrinos are formed from the decay of superheavy dark matter particles and how photons are produced through a new mediator one-loop scattering process. The introduction of photons into the plasma can significantly influence its thermal and ionization evolution history. In this section, we will examine the energy injection rate into the plasma resulting from the radiative scattering of UHE neutrinos with background neutrinos. The cosmic neutrino background temperature evolves as 
\begin{eqnarray}
    T_{\nu_{\mu}}= T_{\nu_{\mu},0}~(1+z)\,.
\end{eqnarray}
The present-day temperature of neutrinos is about \( T_{\nu_{\mu},0} \approx 1.95\, \text{K} \approx 1.7 \times 10^{-4}\, \text{eV} \). At present, neutrinos can be considered nonrelativistic as \( T_{\nu_{\mu}} \ll m_{\nu_{\mu}} \), with the mass of the muon neutrino \( m_{\nu_{\mu}} \sim 0.1\, \text{eV} \). However, at a redshift of \( z \sim 10^6 \), the neutrino temperature rises to \( T_{\nu_{\mu}} \approx 330\, \text{eV} \), and at \( z \sim 5 \times 10^4 \), it is about \( T_{\nu_{\mu}} \approx 8\, \text{eV} \). At these epochs, since \( T_{\nu_{\mu}} > m_{\nu_{\mu}} \), the background neutrinos must be considered as relativistic. In this scenario, the scattering rate of UHE neutrinos interacting with a relativistic background neutrino can be expressed as follows \cite{Döring_2024, PhysRevD.109.063007}:
\begin{eqnarray} \label{eq:intrate}
 \Gamma_{\nu_\mu} &=&\frac{1}{(2 \pi)^3} \int d^3p_{\nu_{\mu}}  \,f(p_{\nu_{\mu}})\ \sigma(s) \, v_{\rm Mol}\,\nonumber\\
 &\approx& 0.1827~ T_{\nu_{\mu}}^3 ~ \sigma(s) \, v_{\rm Mol}\,,
 \end{eqnarray}
where $f(p_{\nu_{\mu}})=1/[e^{p_{\nu_{\mu}}/T_{\nu_{\mu}}}+1]$ is the momentum distribution of the cosmic neutrino background and $v_{\rm Mol}$ is the M{\o}ller velocity. The M{\o}ller velocity is defined as \cite{GONDOLO1991145},
\begin{eqnarray}
    v_{\rm Mol}= \frac {\sqrt{(p_{\nu_{\mu}}\cdot p_{\nu})^2 - m_{\nu_{\mu}}^2m_{\nu}^2}}{E_{\nu_{\mu}}\cdot E_{\nu}},
\end{eqnarray}
where \( E_{\nu_{\mu}} \), \( p_{\nu_{\mu}} \), and \( m_{\nu_{\mu}} \) represent the energy, momentum, and mass of the cosmic background neutrino, respectively, and \( p_{\nu} \), \( m_{\nu} \), and \( E_{\nu} \) denote the momentum, mass, and energy of the UHE neutrino. The energy of the relativistic neutrinos is given by the equation: 
\begin{eqnarray}
E_{\nu_{\mu}} = \sqrt{(p_{\nu_{\mu}}^2 + m_{\nu_{\mu}}^2)}\,. 
\end{eqnarray}
For the incoming UHE neutrino, the momentum can be approximated as \( |p_\nu| \approx E_\nu \), and for the relativistic background neutrino, we have \( |p_{\nu_{\mu}}| \approx E_{\nu_{\mu}} \) since \( E_{\nu_{\mu}} \gg m_{\nu_{\mu}} \). Under these conditions, the M{\o}ller velocity can be approximated as \( v_{\text{Mol}} = (1 - \cos \theta) \), and the center-of-mass energy can be approximated as \cite{GONDOLO1991145}: 
\begin{eqnarray}
s \approx 2\, p_{\nu_\mu} \cdot p_{\nu} \approx 2 E_{\nu_\mu} E_{\nu} (1 - \cos \theta), 
\end{eqnarray}
where \( \theta \) is the angle between the momenta of the UHE and background neutrinos. The term \( (1 - \cos \theta) \) ranges from \( 0 \) (for \( \theta = 0 \)) to \( 2 \) (for \( \theta = \pi \)). When averaging over angles for an isotropic background, the typical contribution is \( (1 - \cos \theta) = \mathcal{O}(1) \). Therefore, the center-of-mass energy essentially simplifies to: 
\begin{eqnarray} 
s \approx 2\, E_{\nu_\mu} E_{\nu},
\end{eqnarray}
where \( E_{\nu_\mu} \approx 3.151 \, T_{\nu_{\mu}} = 3.151 \, T_{\nu_{\mu},0} (1 + z) \). This indicates that the center-of-mass energy decreases as the redshift \( z \) decreases.
 
Following Ref. \cite{PhysRevD.110.123033}, we consider that UHE neutrinos---resulting from the decay of superheavy dark matter---account for a fraction of dark matter. Therefore, the present-day UHE neutrino density $n_{\nu, 0}$ can be written as $n_{\nu, 0}=f_\chi\, \Omega_\chi\,\rho_c/m_\chi\,$,
where $f_\chi$ represents the fraction of the dark matter in the form of heavy dark matter that has decayed into UHE neutrinos. In this work, we take $f_\chi=0.1$ and $1$. $\Omega_\chi$ is the total present-day dark matter abundance, $m_\chi$ is the mass of dark matter, and $\rho_c$ is the critical density of the Universe. The number density of UHE neutrinos varies as $n_{\nu}=n_{\nu, 0}\,(1+z)^3$. The evolution of the number density of the UHE neutrinos $n_{\nu}$ is given by
\begin{eqnarray}\label{eq:numberdensity}
    \frac{dn_\nu}{dt}&=& n_\nu~\Gamma_{\nu_\mu} \nonumber\\ 
    &=& f_\chi\, \rho_c\, \frac{\Omega_\chi}{m_\chi}\, \Gamma_{\nu_\mu}\, (1+z)^3.
\end{eqnarray}
In the case of dark matter decay into a pair of UHE neutrinos, the energy injection rate can be determined by multiplying \eqref{eq:numberdensity} by $m_\chi$, assuming that nearly the entire rest mass of dark matter is converted into the energy of neutrinos. Using Eq. \eqref{eq:numberdensity}, the energy injection rate per unit volume can be written as,
\begin{eqnarray}\label{eq:enrgyinjection}
\frac{dE}{dV~dt} = f_\chi\, \Omega_\chi\,\rho_c\, \Gamma_{\nu_\mu}\, (1+z)^3\,.
\end{eqnarray}
This energy injection into the plasma can significantly affect its thermal and ionization evolution, potentially distorting the CMB spectrum. In this work, we consider the CMB spectral distortion resulting from the energy injection in the plasma due to ultrahigh-energy neutrinos produced by dark matter decay.}

\section{CMB Spectral Distortion}\label{cmbsd}
The CMB spectral distortion provides a unique window into the thermal history of the early Universe, probing the presence of any new physics.
The CMB spectrum is well-approximated by the black body spectrum \cite{Fixsen:1996nj, Mather:1993ij}. As the Universe expands, the photon wavelengths increase, resulting in a lower temperature of the photons and a shift in the peak frequency of the spectrum. However, the shape of the spectrum remains unchanged solely due to the adiabatic expansion of the Universe--- as the expansion uniformly scales the photon energies, maintaining the thermal distribution. When there is an energy injection into the plasma from any exotic source, the CMB spectrum may be distorted from the ideal black body spectrum. These injections introduce nonthermal photons or alter the ionization state of the plasma, leading to processes like Compton scattering or bremsstrahlung, which modify the photon distribution. Consequently, the CMB spectrum may exhibit deviations, broadly classified into two main types: $y$-type distortion or a $\mu$-type distortion. The type of distortion mainly depends on the epoch of the energy injection. The thermal equilibrium between matter and radiation is maintained when energy is injected into plasma at a redshift of $z > 2 \times 10^6$ \cite{khatri2012beyond, Kunze_2014, PhysRevD.111.043002}. During this epoch, photons and baryons efficiently scatter with one another. Therefore, the photons remain in full thermal equilibrium with the baryons. When energy is injected into plasma by any mechanism, the Compton scattering rapidly redistributes the energy into photons, and the double Compton and bremsstrahlung scattering modify the number density of photons to maintain a black body distribution for photons \cite{khatri2012beyond, Kunze_2014}.

As the Universe expands and cools, reaching redshifts $z < 2 \times 10^6$, elastic Compton scattering remains efficient, but double Compton and bremsstrahlung scatterings become gradually inefficient, resulting in less thermalization of the CMB photon distribution. Therefore, any energy injection into the plasma increases the electron temperature, but the number density of photons remains the same due to elastic Compton scattering. These processes establish a Bose-Einstein distribution for the CMB photons, characterized by an effective nonzero chemical potential, rather than restoring the ideal blackbody spectrum--- known as the $\mu$-type of CMB spectral distortion \cite{khatri2012beyond, PhysRevD.85.103522, Chluba:2012gq, Si:2025vsj, Pajer:2012qep, Kunze:2013uja, Kunze_2014, PhysRevD.111.043002, Si:2026nni}. Such distortions are primarily generated at redshifts in the range $5 \times 10^4 < z < 2 \times 10^6$. 
The evolution of $\mu$ distortion due to the energy injection into {plasma} from ultrahigh-energy neutrinos produced by decay of superheavy dark matter can be expressed as \cite{Kunze_2014, PhysRevD.111.043002}, 
\begin{eqnarray}\label{dmu/dt}
     \frac{\partial \mu}{\partial t}= \frac{1}{3}\times1.4\left[\left(\frac{dE}{dVdt}\right)\, \bigg/\,\rho_\gamma\right] -\frac{\mu}{t_{\rm dc}}\,. \label{dmudt}
\end{eqnarray}
The factor of $1/3$ in the first term of Eq. \eqref{dmu/dt} denotes that spectral distortions arise by only $1/3$ of the total energy injection, while the remaining $2/3$ contributes to enhance the average temperature \cite{Chluba:2012gq, Pajer:2012qep, Kunze:2013uja}. Here, the double Compton scattering timescale is denoted by $t_{\rm dc}$:
\begin{eqnarray}
    t_{\rm dc} =2.06\times 10^{33}\left(\frac{1}{\Omega_b h^2}\right)\left(1-\frac{Y_p}{2}\right)^{-1}z^{-9/2}\,,
\end{eqnarray}
where $\Omega_b$ is the baryon energy density parameter. The evolution of $\mu$ with redshift $z$ can be written as \cite{Kunze:2013uja},
 \begin{eqnarray}
     \mu= \frac{1.4}{3}\int  e^{-\left({z}/{z_{\rm dc}}\right)^{5/2}}\ \left[\left(\frac{dE}{dVdz}\right)\, \bigg/\, {\rho_\gamma}\right]\ dz\,,\label{dmudz}
 \end{eqnarray}
where $[{dE}/({dVdz})]=-[{dE}/({dVdt})]\big/[H\,(1+z)]$ represents the energy injection rate per unit volume--- given by equation \eqref{eq:enrgyinjection}. $\Omega_b$ is the baryon energy density parameter. The evolution of $\mu$ with redshift $z$ can be written as in \cite{Kunze:2013uja};the integration limits are from $z=2\times 10^6$ to $z=5\times 10^4$. Furthermore, $\rho_\gamma$ is the photon energy density, and 
\begin{eqnarray}
z_{\rm dc}\equiv 1.1 1\times 10^7\left(\frac{\Omega_bh^2}{0.0224}\right)^{-{2}/{5}}\,,
\end{eqnarray}
where $\Omega_b$ is the dimensionless energy density parameter for the baryons. The COBE/FIRAS experiment placed a 95\% confidence level upper bound on the $\mu$-type spectral distortion to be $\mu < 9.0\times 10^{-5}$ \cite{Fixsen:1996nj}. The upcoming PIXIE experiment is designed to achieve a sensitivity at a level of $\mu = 5 \times 10^{-8}$ at 5$\sigma$ confidence, surpassing current COBE/FIRAS constraints by about 3 orders of magnitude \cite{Kogut:2011xw}.

The $y$-type CMB spectral distortion, known as the Sunyaev-Zel'dovich (SZ) effect, becomes significant when the Universe can no longer fully thermalize injected energy. At lower redshifts ($z < 5 \times 10^4$), the elastic Compton scattering also becomes ineffective in addition to double Compton scattering and bremsstrahlung, causing inefficient thermalization \cite{Sunyaev:1970er}. This type of spectral distortion occurs when hotter electrons transfer their energy to the low-energy CMB photons via the nonrelativistic inverse Compton scattering. The energy transfer causes upscattering of low-frequency photons in the Rayleigh-Jeans (RJ) tail to higher frequencies, reducing intensity in the RJ region. The upscattered photons populate higher frequencies, boosting the intensity in the Wien tail. This results in a characteristic $y$-type of CMB spectral distortion and is parametrized by the Compton $y$ parameter \cite{Kunze:2013uja, khatri2012beyond}. This type of distortion is mainly associated with astrophysical processes like galaxy cluster heating, supernovae, or active galactic nuclei in the late-time Universe. However, the energy injection from various nonstandard physics in the early Universe, such as dark matter annihilation and decay, evaporating primordial black holes, decay of superconducting cosmic string, dark matter viscosity, and primordial magnetic fields, can cause this type of CMB spectral distortion \cite{Acharya:2019xla, PhysRevD.111.043002, Wagstaff:2015jaa, Uchida:2024iog}. 
The evolution of $y$-type CMB spectral distortion due to the energy injection into the plasma by radiative interaction between ultrahigh-energy neutrinos and C$\nu$B can be expressed as follows \cite{Kunze_2014, PhysRevD.111.043002}:
\begin{eqnarray}
    \frac{\partial y}{\partial t}=\frac{1}{12}\left[\left(\frac{dE}{dVdt}\right)\, \bigg/\,{\rho_\gamma}\right]\,.
\end{eqnarray}
Now, the evolution of the $y$ parameter with redshift z can be written as,
 \begin{eqnarray}\label{dydz}
y= \frac{1}{12} \int \ \left[\left(\frac{dE}{dVdz}\right)\, \bigg/\,{\rho_\gamma}\right]\ dz\,.
\end{eqnarray}
The COBE/FIRAS set upper bounds on the \textit{y}-type distortion to be $y = 1.5\times10^{-5}$ with 95\% confidence level \cite{Fixsen:1996nj}, and the future experiment PIXIE expected to detect $y$-type distortion at levels of $y = 10^{-8}$ at a $5\sigma$ level \cite{Kogut:2011xw}.

\section{Result and Discussion} \label{Result}

\begin{figure*}
    \begin{center}
        
        \subfloat[] {\includegraphics[width=3.5in,height=2.5in]{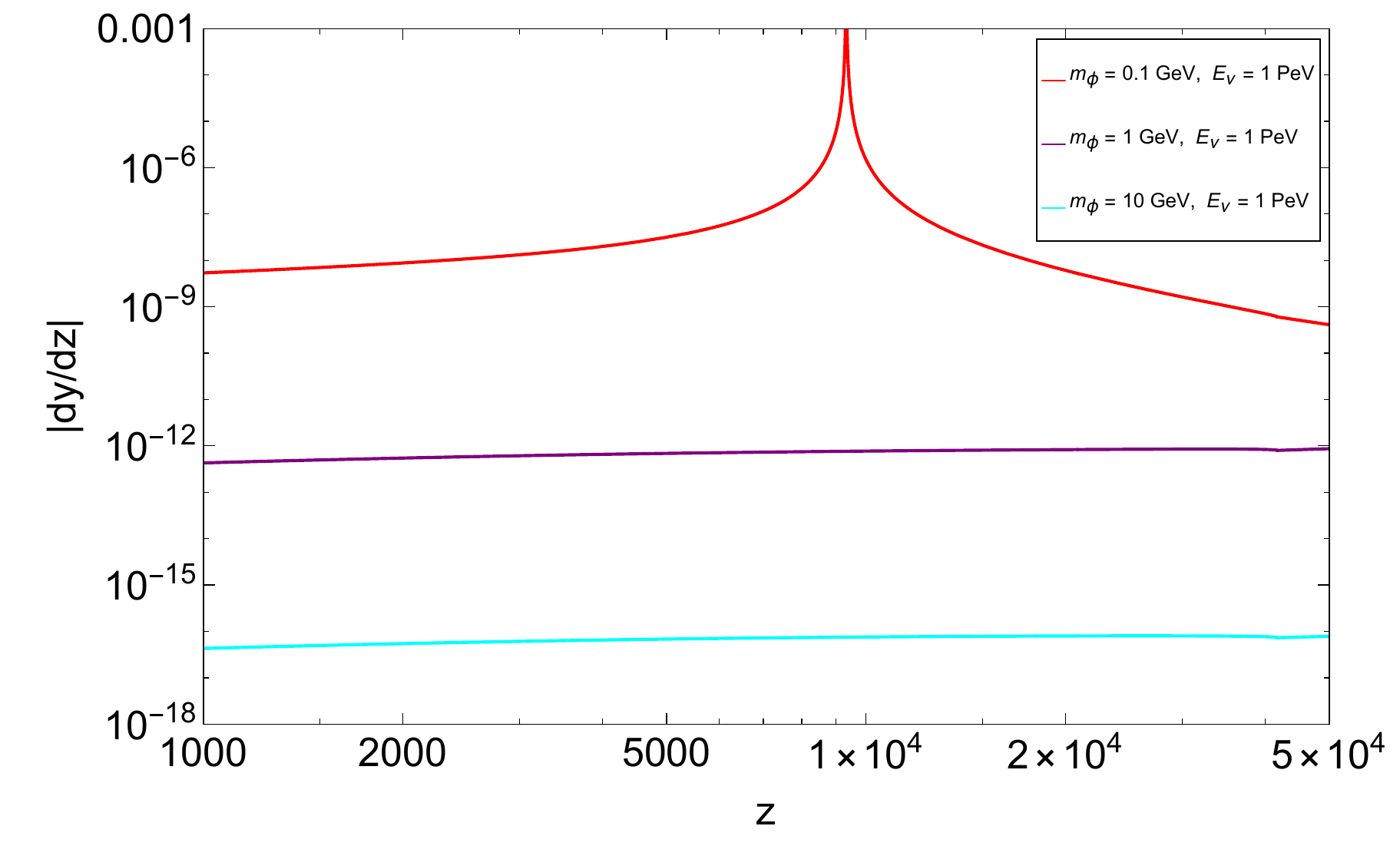}\label{y_evolution}}
        \subfloat[] {\includegraphics[width=3.5in,height=2.5in]{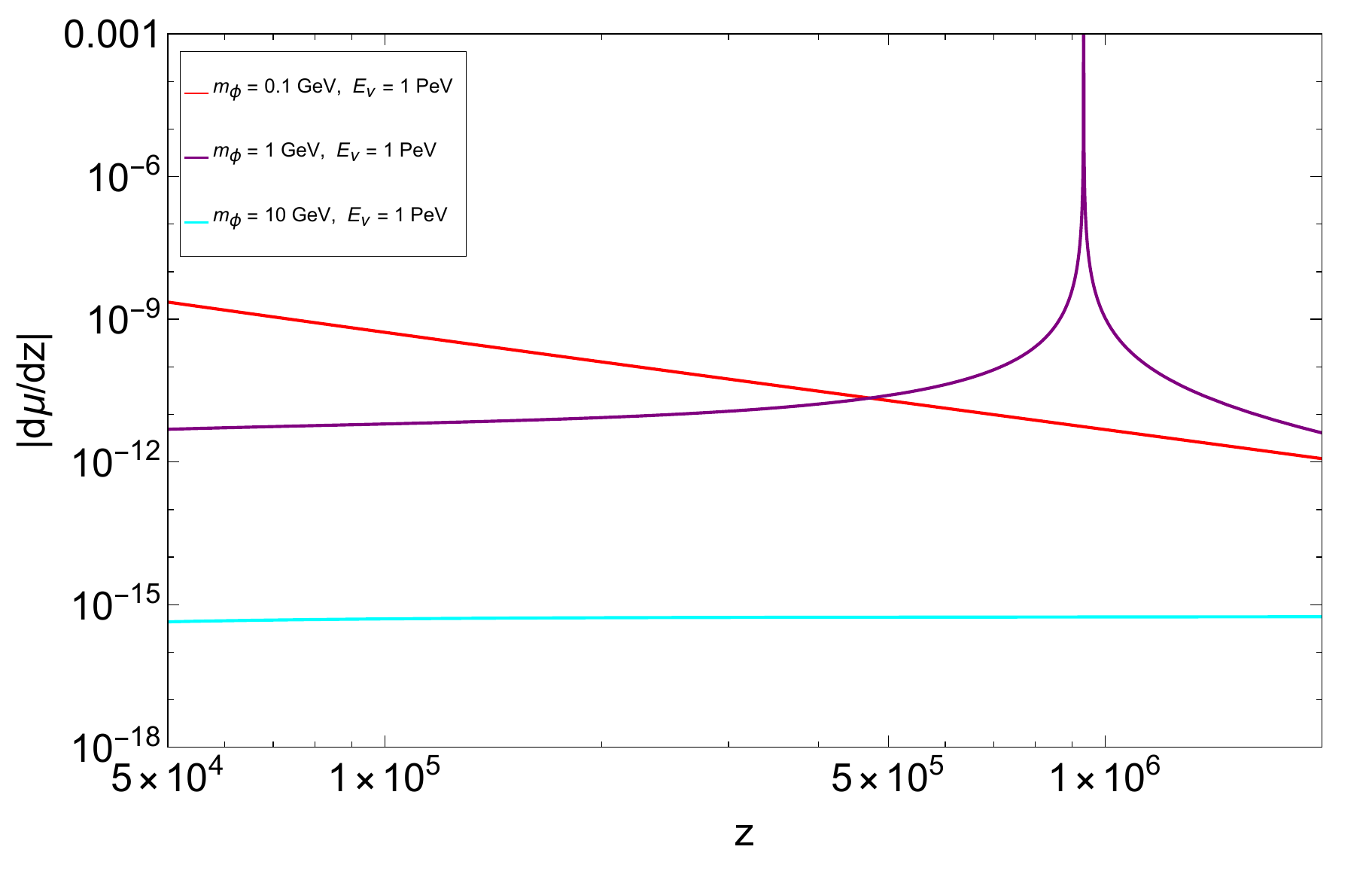}\label{mu_evolution}}     
    \end{center}
    \caption{Evolution of CMB spectral distortions, $|d\mu/dz|$ (left panel), and $|dy/dz|$ (right panel), as a function of redshift $z$ is shown for the case $g_{\nu_\mu}=g_\nu=10^{-2}$ and the UHE neutrino energy to be $E_{\nu}= 1\, \mathrm{PeV}$. The red, purple, and cyan solid coloured lines represent the evolution of  $|dy/dz|$ (left panel) and $|d\mu/dz|$ (right panel), for $f_\chi=0.1$ corresponding to the mediator mass  $m_\phi= 0.1,\, 1,\, {\rm and ~} 10~ \mathrm{GeV}$, respectively.}
    \label{plot: mu_y-evolution}
\end{figure*}

\begin{figure*}
    \begin{center}
        \subfloat[] {\includegraphics[width=3.5in,height=2.5in]{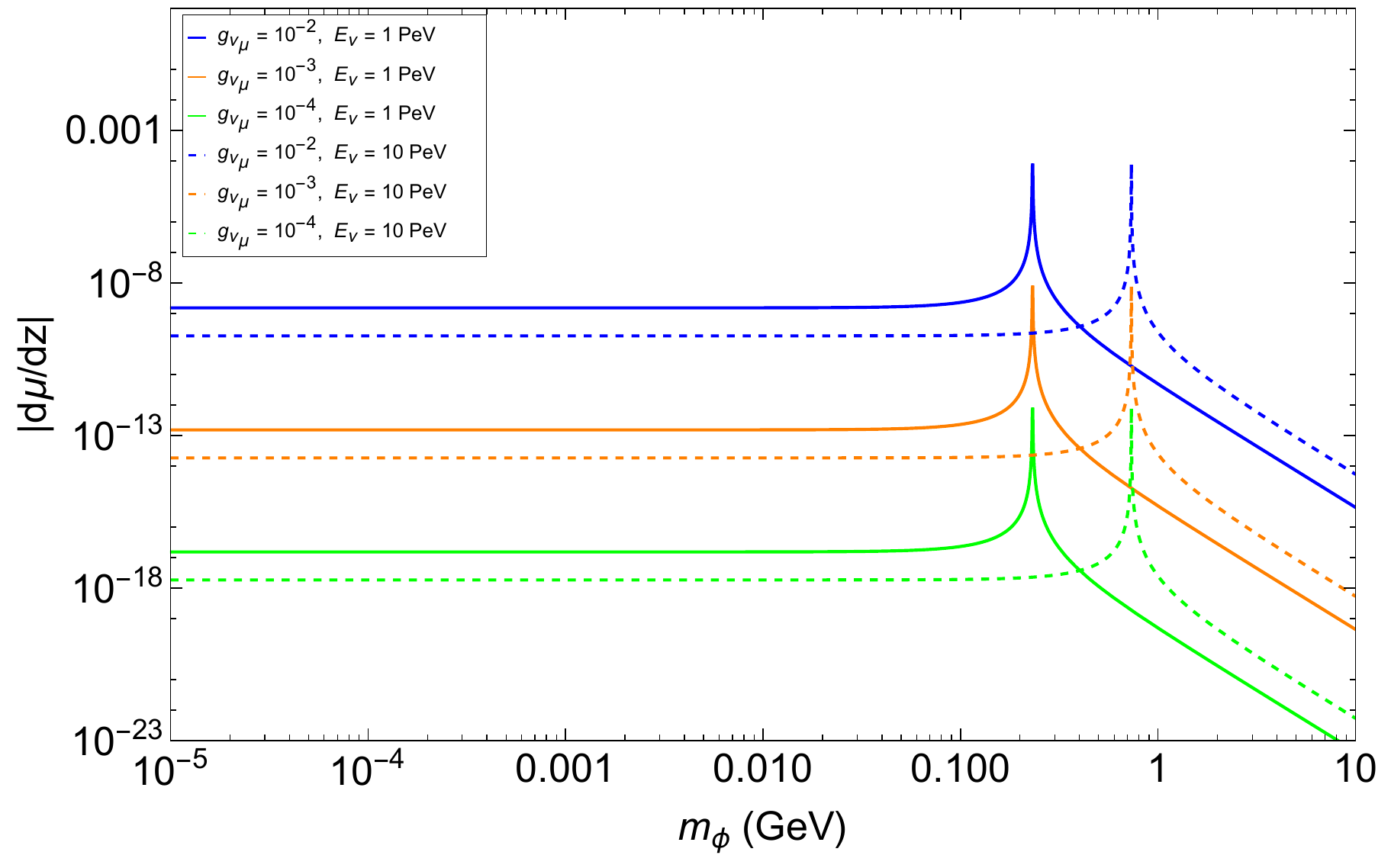}\label{dmu_mphi}}
        \subfloat[] {\includegraphics[width=3.5in,height=2.5in]{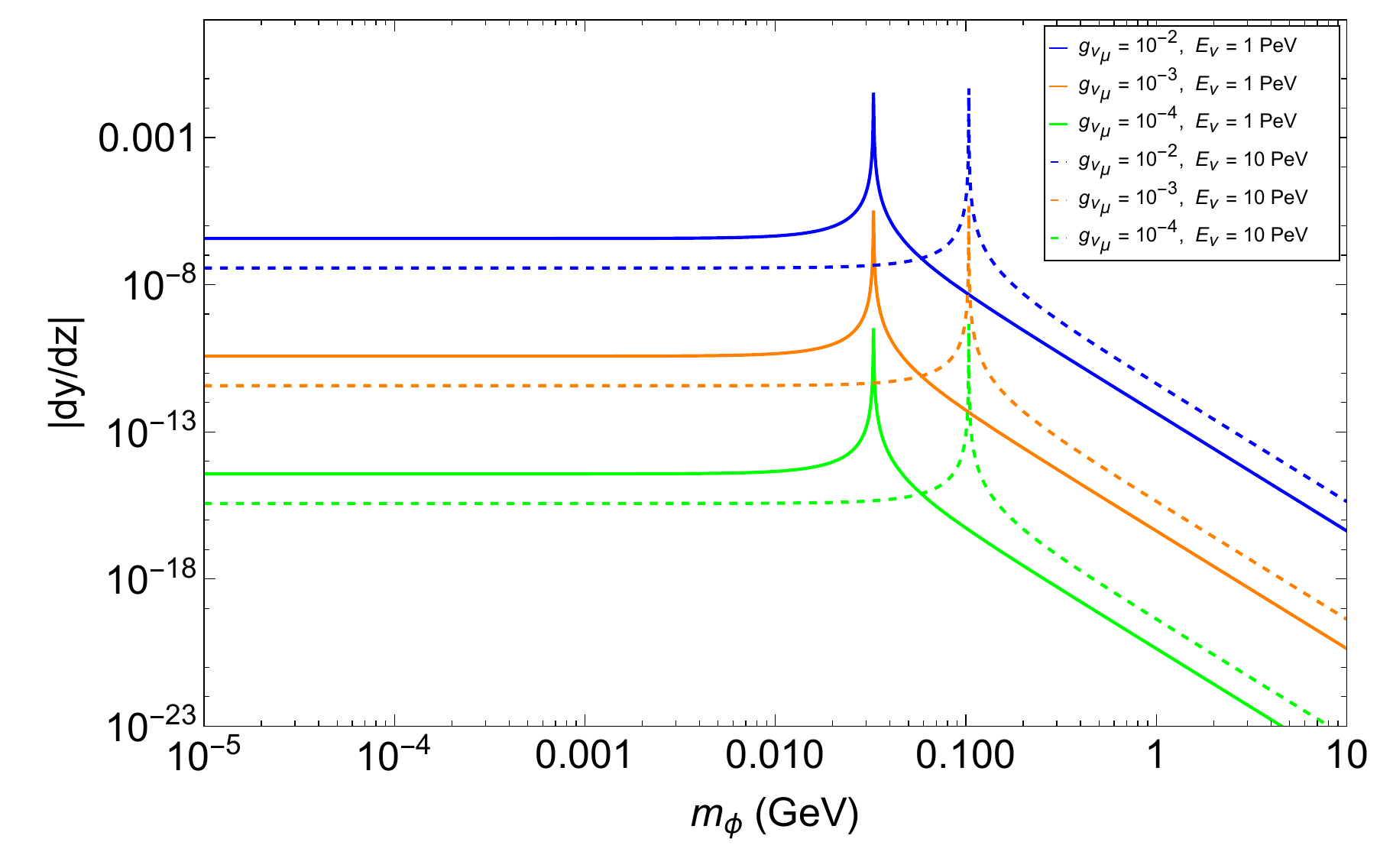}\label{dy_mphi}}    
    \end{center}
    \caption{Evolution of CMB spectral distortions as a function of the mass of mediator ($m_\phi$) is shown in the presence of radiative scattering of ultrahigh-energy neutrinos with cosmic neutrino background \((C\nu B)\). In the left panel, we show $\mu$ type distortion at a redshift of $z=5\times10^{4}$; where, in the right panel, we show $y$ type distortion at a redshift of $z=10^{3}$. In both plots, the solid and dashed coloured lines represent the cases when $E_{\nu } = 1~{\rm PeV}$ and $10\,\ \mathrm{PeV} $, respectively. Here, we vary self-interacting neutrino coupling as $g_{\nu_\mu}= 10^{-2},\, 10^{-3},$ and $ 10^{-4}$ with fixed $f_\chi=0.1$.}
    \label{plot:dydz_m}
\end{figure*}



\begin{figure*}
\begin{center}
     \subfloat[] {\includegraphics[width=3.5in,height=2.5in]{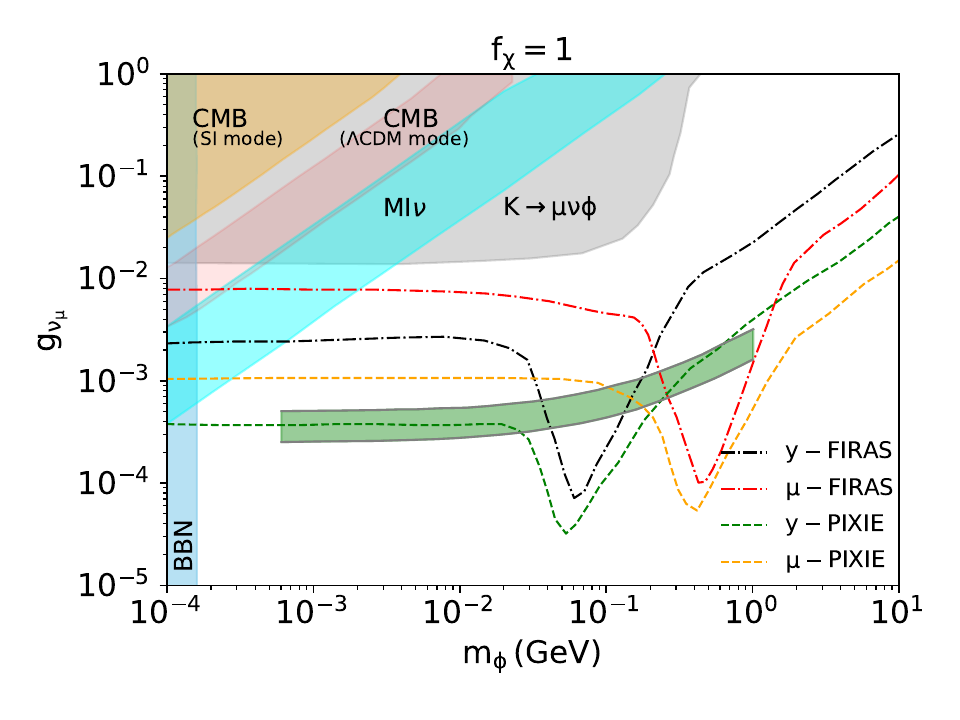}\label{p3a}}
    \subfloat[] {\includegraphics[width=3.5in,height=2.5in]{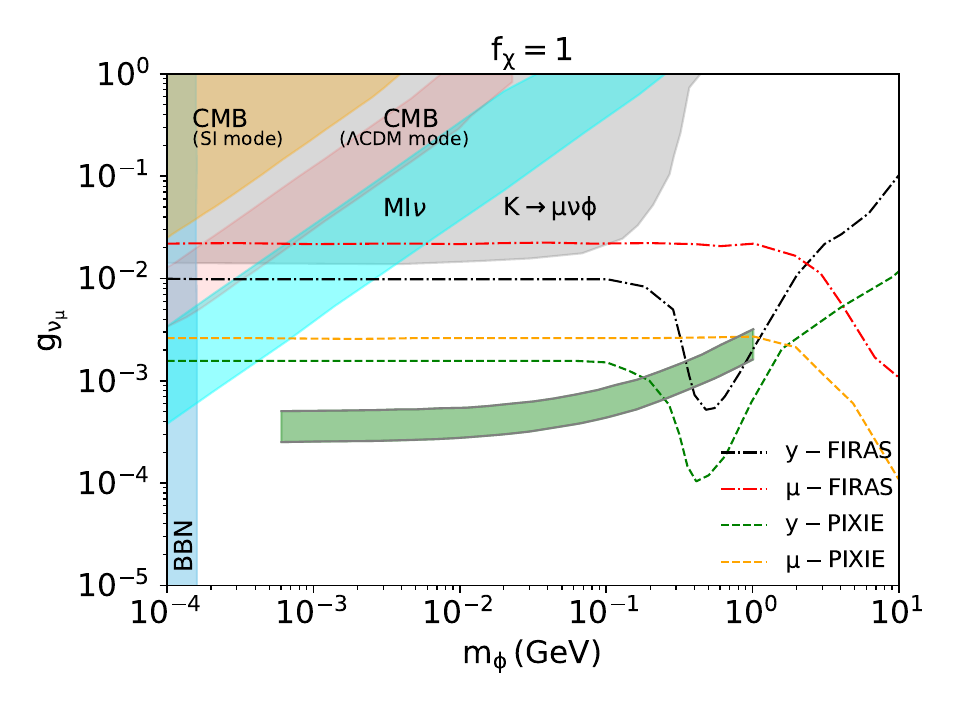}\label{p3b}}\\

     \subfloat[] {\includegraphics[width=3.5in,height=2.5in]{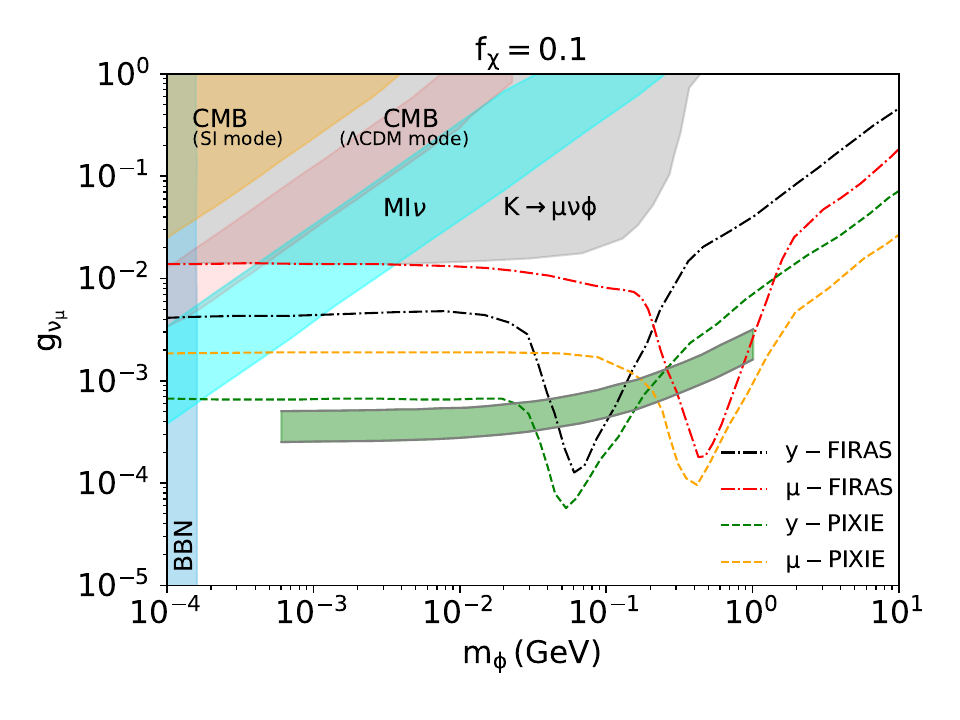}\label{p3c}}
    \subfloat[] {\includegraphics[width=3.5in,height=2.5in]{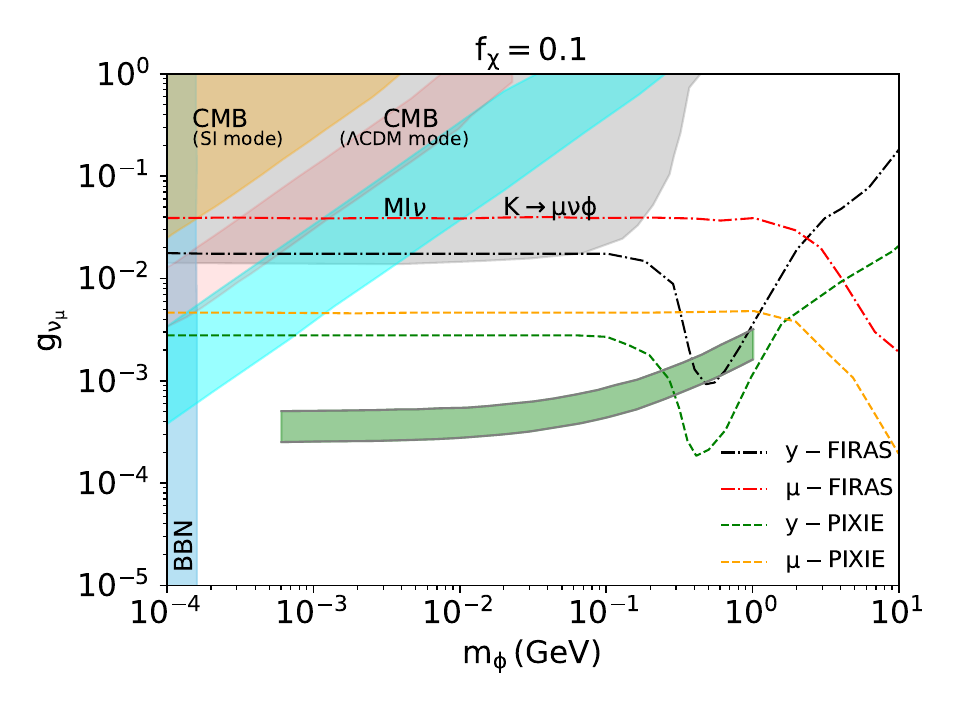}\label{p3d}}
\end{center}
\caption{Upper bounds on the self-interacting neutrino coupling ($g_{\nu_\mu}$) as a function of the mediator mass ($m_\phi$) using PIXIE and FIRAS upper constraints on $\mu$ and $y$ types of CMB spectral distortion are shown for the case $g_{\nu_\mu}=g_\mu$, for the UHE neutrinos energies to be $E_\nu= 1\ \rm{PeV}$ (left panels) and $E_\nu= 100\ \rm{PeV}$ (right panels). Upper panels are obtained for $f_\chi=1$ and lower panels are obtained for $f_\chi=0.1$. { The sky-blue-shaded region denotes the parameter space excluded by the BBN \cite{Huang:2017egl}. The yellow and magenta-shaded regions depict the exclusion from CMB anisotropy measurements \cite{Barenboim:2019tux}.} The gray-shaded region represents the excluded parameter space from the bound on the branching ratio of kaon decay--- $K\rightarrow \mu \nu\phi$ \cite{PhysRevLett.124.041802}. The cyan-shaded region corresponds to the region that can explain the Hubble tension considering moderately interacting neutrino ($\mathrm{MI}\nu$) \cite{Lancaster_2017, PhysRevD.109.063007}, and the green-shaded region can explain the observed $(g-2)_\mu$ anomaly \cite{PhysRevD.105.L051702, PhysRevLett.126.141801}.}
\label{plot:contour}
\end{figure*}

\begin{figure*}
\begin{center}
    \subfloat[] {\includegraphics[width=3.5in,height=2.5in]{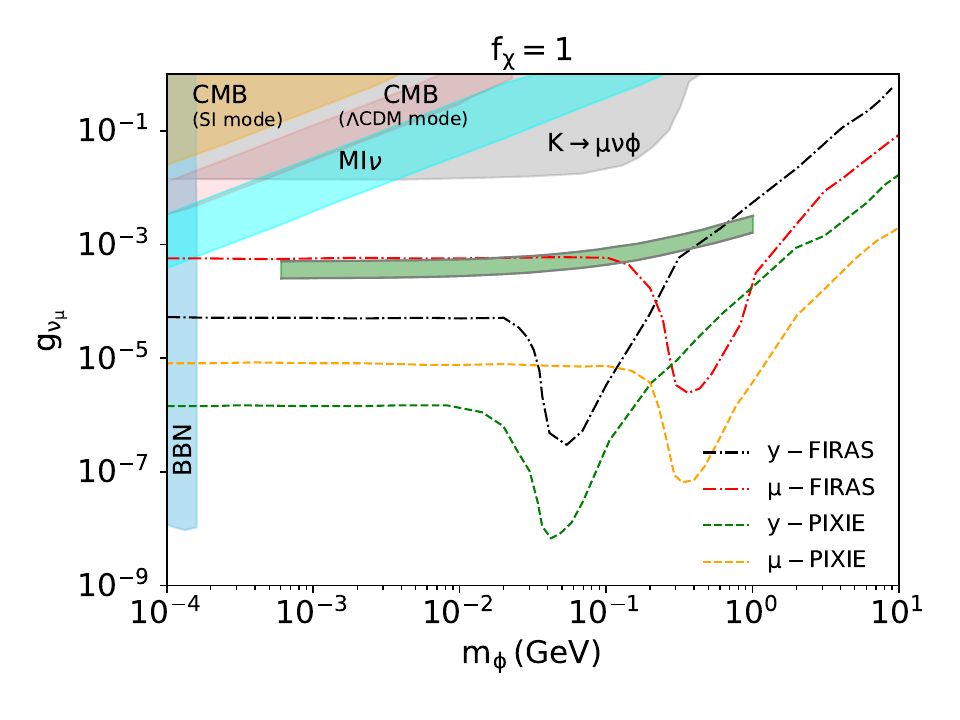}\label{p4a}}
    \subfloat[] {\includegraphics[width=3.5in,height=2.5in]{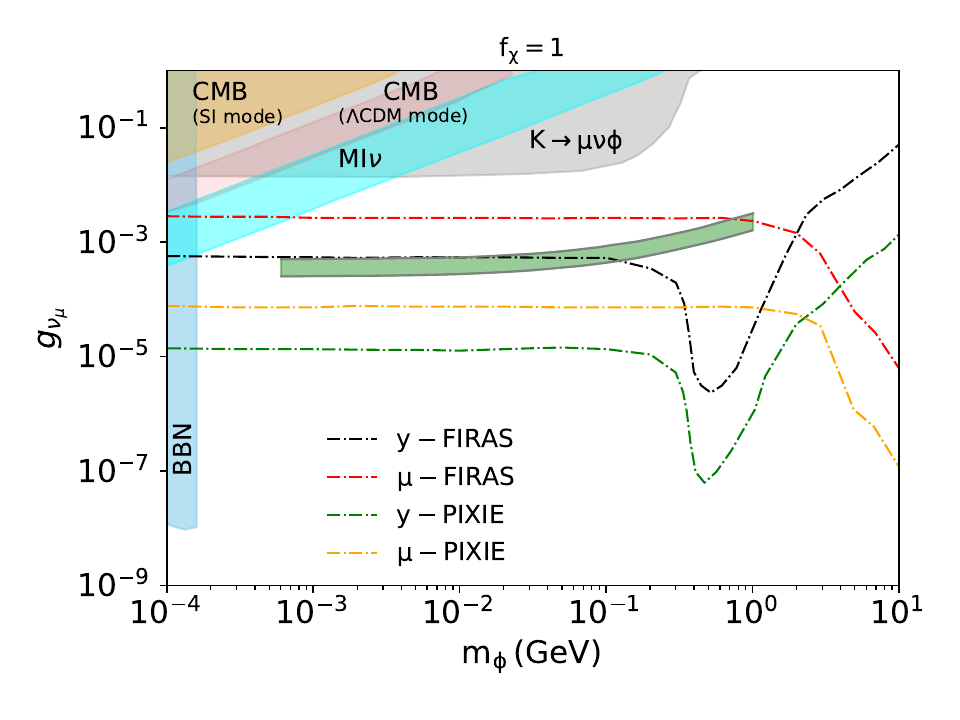}\label{p4b}}\\

    \subfloat[] {\includegraphics[width=3.5in,height=2.5in]{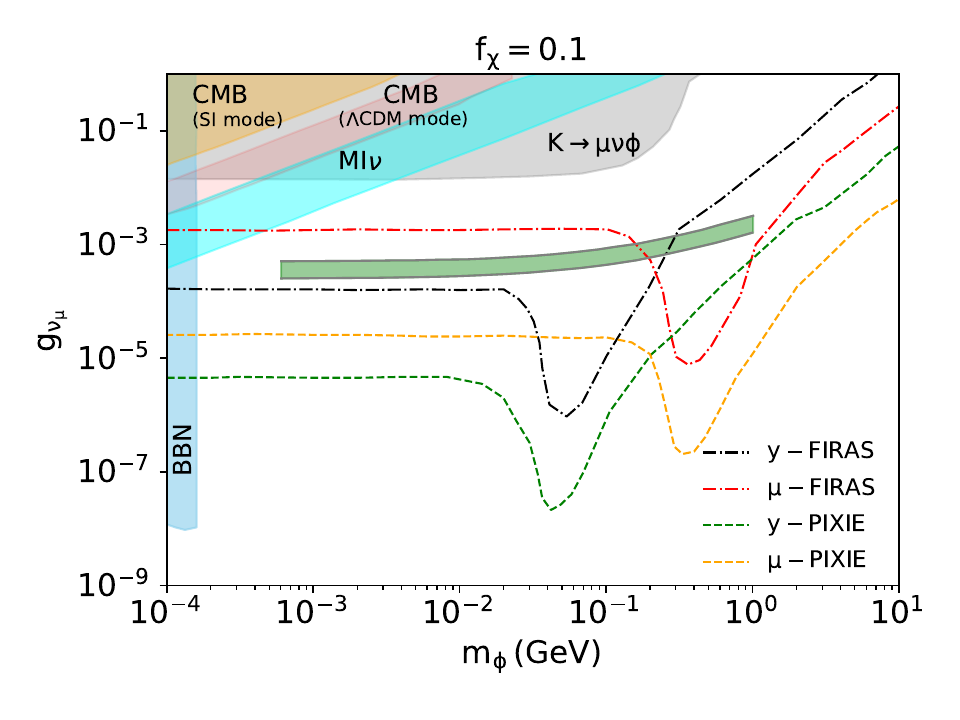}\label{p4c}}
    \subfloat[] {\includegraphics[width=3.5in,height=2.5in]{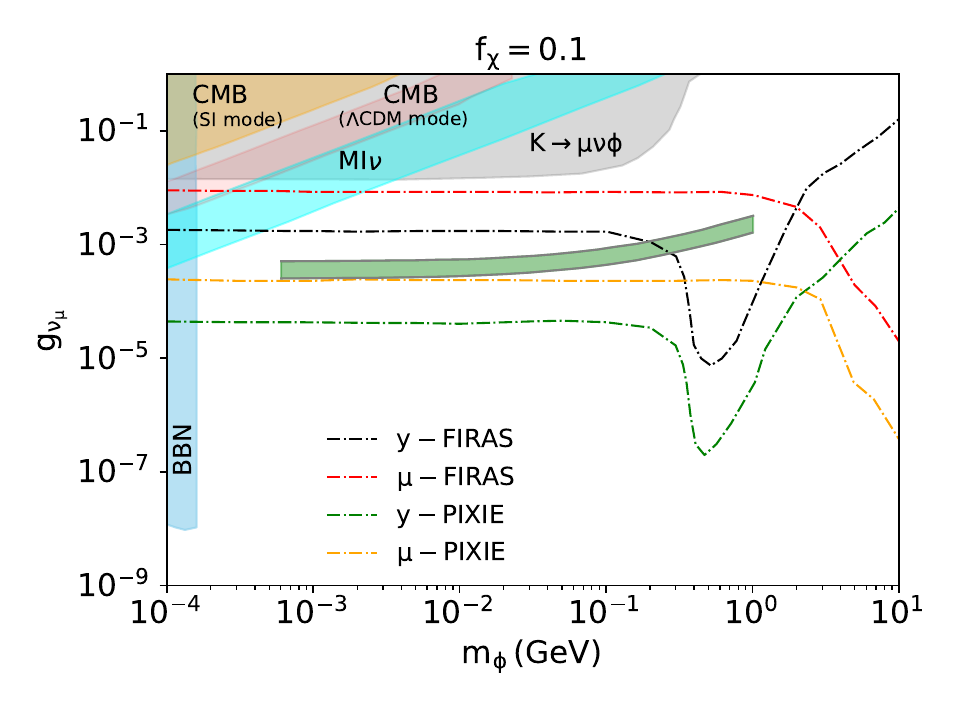}\label{p4d}}
    
\end{center}
\caption{Upper bounds on the self-interacting neutrino coupling ($g_{\nu_\mu}$) as a function of the mediator mass ($m_\phi$) from $\mu$ and $y$ types of CMB spectral distortion for the case $g_{\nu_\mu}\neq g_\mu$. Upper panels are obtained for $f_\chi=1$ and lower panels are obtained for $f_\chi=0.1$. Here, we vary the neutrino energies $E_\nu= 1\ \rm{PeV}$ (left panel) and $E_\nu= 100\ \rm{PeV}$ (right panel) while keeping $g_\mu=0.1$. { We have also plotted existing bounds from BBN (sky-blue-shaded), CMB (yellow and magenta-shaded), from kaon decay (grey-shaded), and the solution to Hubble tension from MI$\nu$ (cyan-shaded).}}
\label{plot:contour2}
\end{figure*}

\begin{figure*}
\begin{center} 
     \subfloat[] {\includegraphics[width=3.5in,height=2.5in]{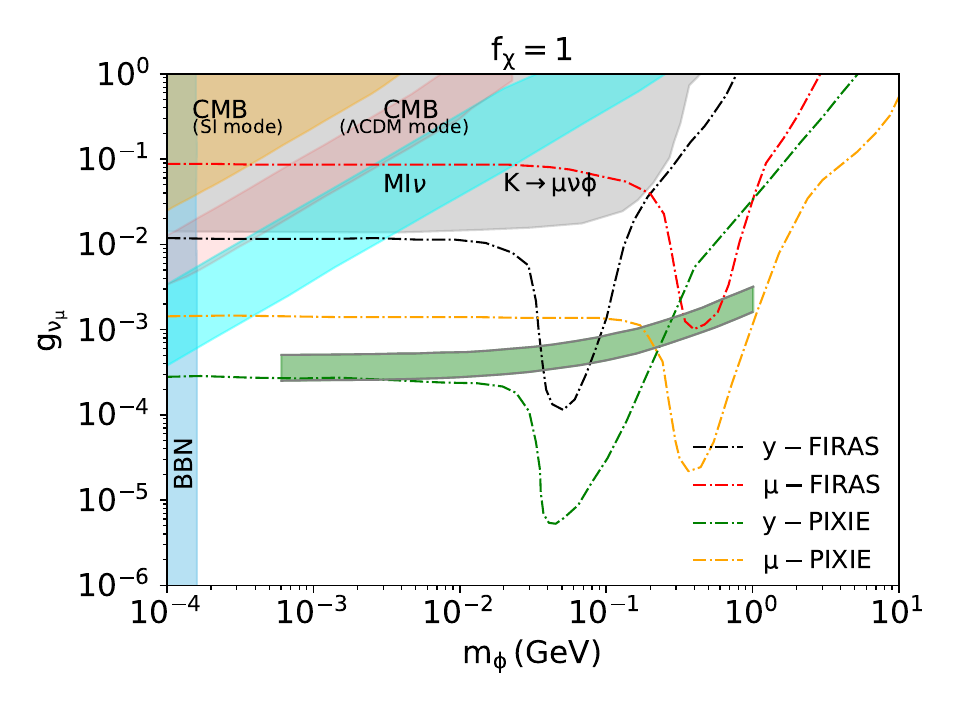}\label{p5a}}
    \subfloat[] {\includegraphics[width=3.5in,height=2.5in]{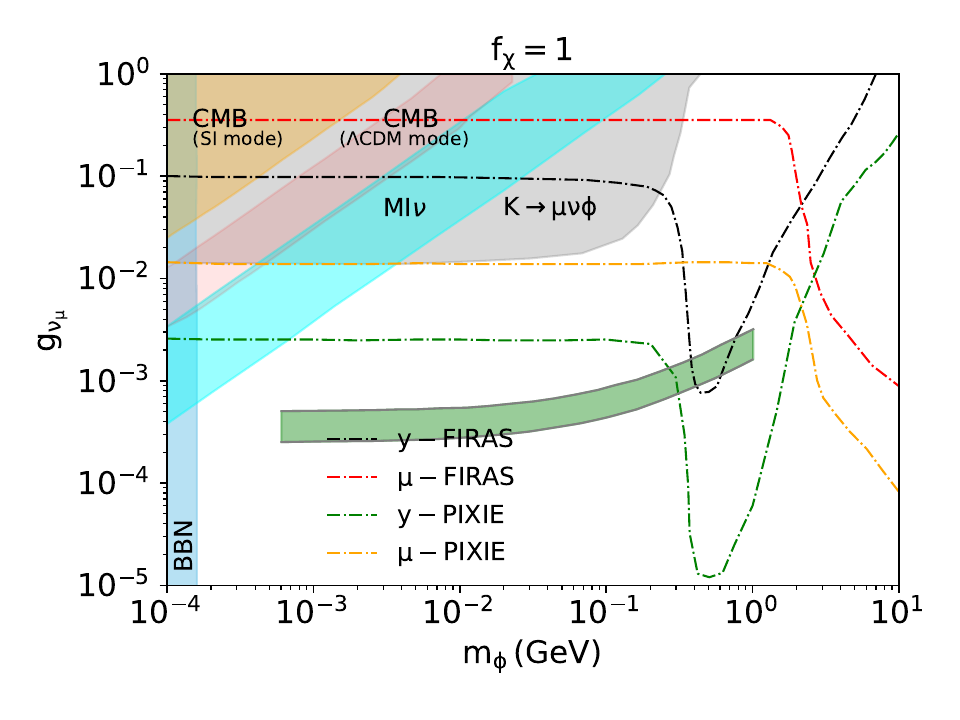}\label{p5b}}\\

    \subfloat[] {\includegraphics[width=3.5in,height=2.5in]{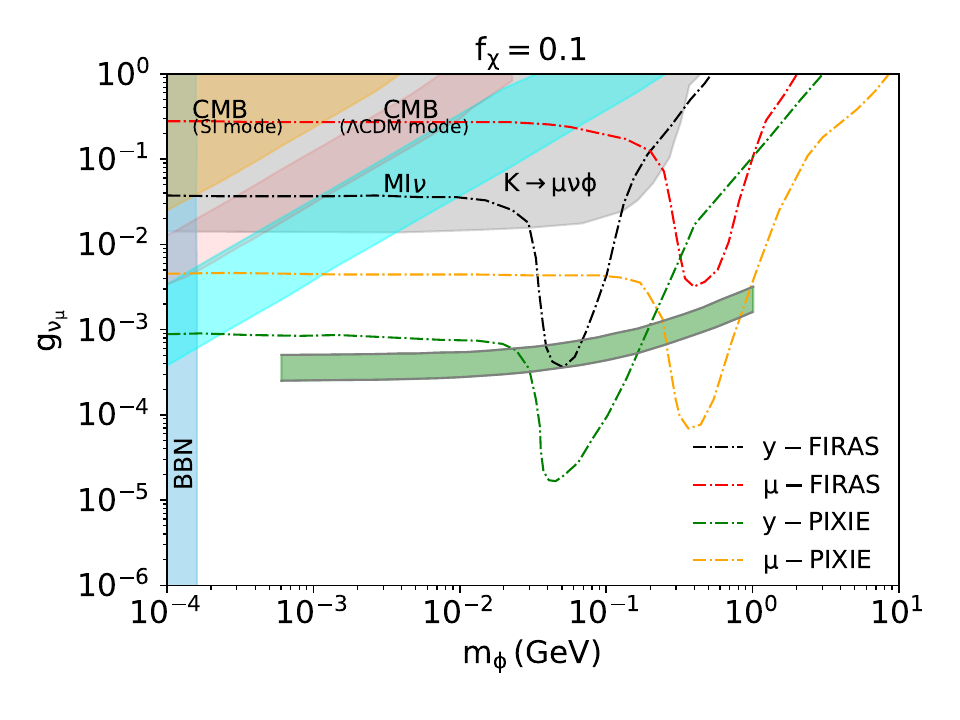}\label{p5c}}
    \subfloat[] {\includegraphics[width=3.5in,height=2.5in]{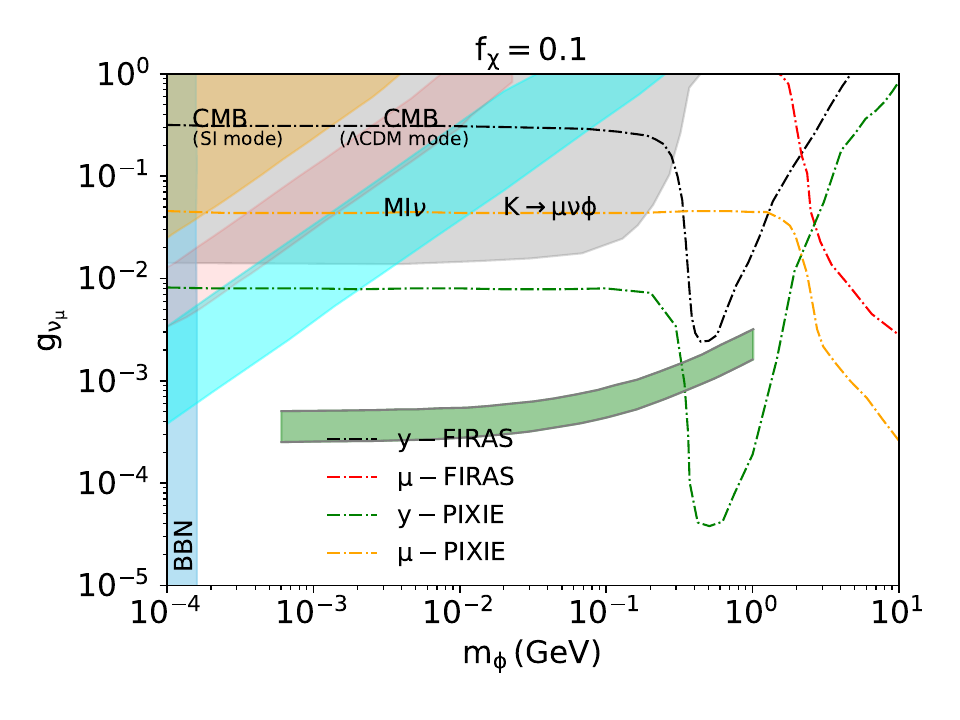}\label{p5d}}
\end{center}
\caption{Upper bounds on the self-interacting neutrino coupling $g_{\nu_\mu}$ as a function of the mediator mass $m_\phi$ from $\mu$ and $y$ types of CMB spectral distortion for the case $g_{\nu_\mu}\neq g_\mu$. Upper panels are obtained for $f_\chi=0.1$ and lower panels are obtained for $f_\chi=1$. Here, we vary the neutrino energies $E_\nu= 1\ \rm{PeV}$ (left panel) and $E_\nu= 100\ \rm{PeV}$ (right panel) for $g_\mu=5\times10^{-4}$ \cite{PhysRevD.105.L051702}. { We have also shown existing bounds from BBN (sky-blue-shaded), CMB (yellow and magenta-shaded), from kaon decay (grey-shaded), and the solution to Hubble tension from MI$\nu$ (cyan-shaded).}}
\label{plot:contour3}
\end{figure*}

In this section, we present our numerical analysis of the $\mu$ and $y$ types of distortion, focusing on the radiative scattering of self-interacting ultrahigh-energy neutrinos into photons. First, we analyze how significantly energy injection due to the radiative scattering affects the evolution of the $\mu$ and $y$ types of CMB distortions. In Figs. (\ref{plot:dydz_m}) and (\ref{plot: mu_y-evolution}), we have plotted the evolution of these distortions as a function of redshift ($z$) and the mass of the mediator ($m_\phi$).

{ In Fig. \ref{plot: mu_y-evolution}, we illustrate the evolution of CMB spectral distortions as a function of redshift for a fixed ultrahigh-energy neutrino energy of \(E_\nu = 1 \, \text{PeV}\). In Figures \ref{plot: mu_y-evolution} and \ref{plot:dydz_m}, we use \( f_\chi = 0.1 \) for illustrative purposes. Figure \ref{y_evolution} shows the evolution of \(y\)-type distortions over the redshift range from \(z = 5 \times 10^4\) to \(z = 10^3\). In Figure \ref{mu_evolution}, we plot the evolution of \(\mu\)-type distortions for a redshift range of \(z = 2 \times 10^6\) to \(z = 5 \times 10^4\). 

In Figure \ref{y_evolution}, a notable kink occurs for \(m_\phi = 0.1 \, \text{GeV}\) at a redshift of approximately \(9.2 \times 10^3\). Similarly, in Figure \ref{mu_evolution}, the kink appears for \(m_\phi = 1 \, \text{GeV}\) at a redshift of around \(9.2 \times 10^5\). These kinks arise because the energy injection rate is directly proportional to the cross-section, as indicated in Equations \eqref{eq:intrate} and \eqref{eq:enrgyinjection}. 

The cross-section is roughly proportional to the expression \( {s}/{\left[(m_\phi^2 \, g_\mu \, g_{\nu_\mu}/4\pi)^2 + s^2(1 - m_\phi^2/s)^2\right]}\). When the center-of-mass energy \(\sqrt{s} < m_\phi\) or \(\sqrt{s} > m_\phi\), the denominator term \(s^2(1 - m_\phi^2/s)^2\) dominates, particularly because \((g_\mu \, g_{\nu_\mu}/4\pi)^2 \ll 1\). Here, taking \(g_\mu = g_{\nu_\mu} = 10^{-2}\) gives us \((g_\mu \, g_{\nu_\mu}/4\pi)^2 \approx \mathcal{O}(10^{-10})\), which is negligible compared to the term \(s^2(1 - m_\phi^2/s)^2\). 

As the center-of-mass energy approaches the mediator mass (\(\sqrt{s} \rightarrow m_\phi\)), the term \(m_\phi^2\Gamma_\phi^2\) then dominates over \((s - m_\phi^2)\) in the denominator of the cross-section, according to Equation \eqref{eq:crossec}. This results in exceptionally high values of the cross-section compared to the previous two cases where \(\sqrt{s} < m_\phi\) or \(\sqrt{s} > m_\phi\). The significant increase in the cross-section leads to the observed kink in distortion, as shown in Figure \ref{plot: mu_y-evolution}.

As previously discussed in Section \ref{energy-injection}, the center-of-mass energy is proportional to redshift (\(s \propto (1 + z)\)). Therefore, for a higher mediator mass, the resonance condition \(\sqrt{s} \approx m_\phi\) is met at a higher redshift, causing the corresponding kink to appear earlier in cosmic time.

The $\mu$-type distortion happens in the radiation-dominated era. The photon number density decreases with decreasing redshift, due to the expansion of the Universe, as described by the relation, $\rho_\gamma \propto (1+z)^4$. The energy injection into the plasma follows the relation, $[dE/(dV\,dz)]\propto (1+z)^5~\sigma(s)/H(z)$, as seen in Eqs. \eqref{eq:intrate} and \eqref{eq:enrgyinjection}. During the radiation-dominated era, \({dE}/(dV\,dz)\) behaves like \(\propto (1+z)^3\,\sigma(s)\), while in the matter-dominated era, it behaves like \(\propto (1+z)^{3.5}\,\sigma(s)\). For $\sqrt{s}<m_\phi$, the cross-section is proportional to $s/m_\phi^4\propto(1+z)$ while for $\sqrt{s}>m_\phi$, the cross-section is $\propto 1/s\propto1/(1+z)$. Consequently, when $\sqrt{s}<m_\phi$, the quantity $|d\mu/dz|$ becomes nearly constant (see cyan colour line in Fig.~\ref{mu_evolution}). In contrast, for $\sqrt{s}>m_\phi$, $|d\mu/dz|$ nearly follows the redshift as: $\propto 1/(1+z)^{2}$ --- Eq. \eqref{dmudt} (see red colour line in Fig. \ref{mu_evolution}). After matter-radiation equality ($z\sim3400$), $|dy/dz|$ follows the redshift as $\propto (1+z)^{1/2}$--- i.e. the value of $|dy/dz|$ decreases with decreasing redshift. This pattern can be seen in Fig. \ref{y_evolution} toward the redshift end. 
In both figures, we present variations in distortions for three different values of the mediator mass, \(m_\phi\): \(0.1,\, 1\), and \(10\, \mathrm{GeV}\). As we increase the mediator mass from \(0.1\, \mathrm{GeV}\) to \(10\, \mathrm{GeV}\), both the $\mu$-type and \(y\)-type spectral distortions decrease, as indicated by the red solid lines and the cyan solid lines. It happens because, for a higher mediator mass, the interaction cross-section is suppressed in the regime $\sqrt{s}<m_\phi$, leading to a reduced energy-injection rate. Consequently, a smaller cross-section results in lower spectral distortion.}

In figures \ref{dmu_mphi} and \ref{dy_mphi}, we show the redshift derivative of $\mu$-type and $y$-type of CMB spectral distortions, respectively, as a function of the mass of mediator $m_\phi$ for different neutrino coupling strengths ($g_{\nu_\mu}$) and different energies of UHE neutrinos. In Fig. (\ref{dmu_mphi}), we show $|d\mu/dz|$ at a redshift of $z=5\times10^{4}$--- i.e., the lower limit for $\mu$-type distortion, below which elastic Compton scattering can no longer establish a Bose-Einstein spectrum and energy injection instead produces $y$-type distortions. In Fig. (\ref{dy_mphi}), we show $|dy/dz|$ at redshift a redshift of $z=10^{3}$--- about the redshift of recombination. We vary the mass of the mediator ($m_\phi$) from $10^{-5}$~GeV to 10~GeV. In both plots, the distortion remains constant for lower mediator mass, but it starts to decrease as we go toward higher values of mediator mass after a certain value of \(m_\phi\). It happens because the energy injection rate is directly proportional to the cross-section $dE/dVdt \propto \sigma$; the cross-section is roughly $\propto s/(s^2\,[(m_\phi^2/s\,\times \,g_\mu\,g_{\nu_\mu}/4\pi)^2+(1-m_\phi^2/s)^2])$. When mediator mass is smaller than the centre-of-mass energy, $m_\phi<\sqrt{s}$, the cross-section becomes nearly constant, $\propto s/s^2$, as the factor $(g_\mu\,g_{\nu_\mu}/4\pi)^2$ is already $\ll1$ and $s$ is constant for a fixed redshift. This results in a constant energy injection rate, which leads to a constant spectral distortion of the CMB. When the mediator mass is larger than the center-of-mass energy, $m_\phi>\sqrt{s}$, the cross-section becomes nearly $\propto 1/m_\phi^4$, leading to a reduction in distortion as \(m_\phi\) increases. 
When mediator mass reaches the centre-of-mass energy, $m_\phi\rightarrow\sqrt{s}$, the term $m_\phi^2\Gamma_\phi^2$ dominates over $(s-m_\phi^2)$ in the denominator in cross-section--- equation  \eqref{eq:crossec}. Therefore, the cross-section roughly becomes $\sigma\sim(81\,\alpha^2/4\pi^3)\times(4\pi/m_\phi)^2$. This results in a very high values of the cross-section compared to earlier two cases when $m_\phi<\sqrt{s}$ or $m_\phi>\sqrt{s}$ as in these two cases, the cross-section was multiplied with $(g_\mu\,g_{\nu_\mu})^2$, and $g_\mu\,g_{\nu_\mu}\ll 1$. The very high value of the cross-section results in a kink in distortion, as seen in Fig. \ref{plot:dydz_m}. If we increase the energy of ultrahigh-energy neutrinos, $E_\nu$, the center-of-mass energy $\sqrt{s}$ increases, shifting the kink position towards higher mediator mass. 
In both Figs. \ref{dmu_mphi} and \ref{dy_mphi}, we also show the variation of the coupling parameter in the range of $10^{-2}-10^{-4}$ considering $g_{\mu}=g_{\nu_\mu}$. For a change in the coupling parameter by one order, there is about a four-order change in the distortion of the CMB. It happens because the energy injection rate is proportional to the fourth power of $g_{\nu_\mu}$, i.e., $dE/dV/dt \propto \sigma \propto g_{\nu_\mu}^4$, as shown in Eqs. \eqref{eq:crossec} and \eqref{eq:enrgyinjection}. Therefore, decreasing $g_{\nu_\mu}$ from $10^{-2}$ to $10^{-3}$, $|d\mu/dz|$ decreases from $1.5\times10^{-9}$ to $1.5\times10^{-13}$--- when \(E_\nu=1~{\rm PeV}\). This pattern can also be seen for $|dy/dz|$ in Fig. \ref{dy_mphi}.

In Fig. \ref{plot:contour}, we have constrained the self-interacting coupling strength $g_{\nu_\mu}$ for muon flavor active neutrino as a function of the mediator mass $m_\phi$ for case $g_{\nu_\mu}= g_\mu$ using FIRAS and PIXIE bounds on the $\mu$ and $y$ types of CMB spectral distortions. In Figs. \ref{p3a} and \ref{p3c}, we consider the UHE neutrino energy to be $E_\nu=1~\rm{PeV}$, and in Fig. \ref{p3b} and \ref{p3d}, the UHE neutrino energy is set to 100~PeV. In all Figs. \ref{p3a}, \ref{p3b}, \ref{p3c} and \ref{p3d}, the dashed yellow and green lines illustrate the upper bounds derived from the PIXIE forecast for the $\mu$ and $y$ types of CMB distortions, respectively, corresponding to $5 \times 10^{-8}$ and $10^{-8}$. The red and black dot-dashed lines indicate the constraints obtained using FIRAS bounds on the $\mu$ and $y$ types of CMB distortions, respectively, corresponding to $9 \times 10^{-5}$ and $1.5 \times 10^{-5}$.

We find upper bounds on the self-interacting coupling parameter to be {\(1.0 \times 10^{-3}\) and \(3.8 \times 10^{-4}\)} using forecasted PIXIE limits on the \(\mu\) and \(y\) types of CMB spectral distortion, respectively, while using the FIRAS bound on $\mu$ and $y$ types distortion, the upper bounds on the coupling parameter become {\(8.0 \times 10^{-3}\) and \(2.2 \times 10^{-3}\) for $f_\chi=1$--- as shown in figure \ref{p3a}. Reducing the value of $f_\chi$ by tenfold from 1 to 0.1 relaxes the bounds by approximately a factor of 1.8--- as shown in figure \ref{p3c}. This pattern is consistent for every tenfold change in the value of $f_\chi$.}
Here, we have assumed \(g_{\nu_\mu} = g_\mu\) and the UHE neutrino energy to be 1 PeV. The bounds on the coupling parameter \( g_{\nu_\mu} \) remain constant until the mass of the mediator approaches the center-of-mass energy, \( \sqrt{s} \). This occurs because, at lower values of \( m_\phi \), the energy injection does not depend on the mediator mass, as discussed earlier. When the mediator mass reaches the center-of-mass energy $\sqrt{s}$, a kink appears in all Figs \ref{p3a}, \ref{p3b}, \ref{p3c} and \ref{p3d} as the interaction cross-section attains its maximum value for a given set of parameters. When \( m_\phi > \sqrt{s} \), the upper bound on the coupling parameter becomes nearly proportional to the mediator mass, \( g_{\nu_\mu} \propto m_\phi \). It happens because one needs to keep a constant energy injection rate (eq \eqref{eq:enrgyinjection}) while varying the mass of the mediator so that the value of spectral distortion by energy injection remains less than or equivalent to upper limits on CMB spectral distortion by FIRAS or PIXIE. To keep the rate of the energy injection constant, the value of the cross-section must be fixed (refer to equation \eqref{eq:crossec})--- i.e., the value of \(g_{\nu_\mu}\) needs to be increased proportionally to  \(m_\phi\) since the cross-section follows $\sigma\propto1/m_\phi^4$ for \(m_\phi>\sqrt{s}\), as discussed in figure \ref{plot:dydz_m}. As a result, the limit on the coupling parameter becomes relaxed as we increase the mass of the mediator. In Fig. \ref{p3b} and \ref{p3d}, we derived the upper bounds for UHE neutrinos energy to be $E_\nu= 100\ \rm{PeV}$. For the case where $f_\chi = 1$, we obtain upper bound on the coupling parameter to be {$2.6\times10^{-3}$ and $1.5\times10^{-3}$} using the forecasted PIXIE limit on $\mu$ and $y$ types of distortion, respectively; while using the FIRAS bound on $\mu$ and $y$ types distortion, the upper bounds on the coupling parameter become {$1.0\times10^{-2}$ and $2.1\times10^{-2}$}, respectively. { When $f_\chi=0.1$, the bounds get relaxed by a factor of $~1.8$--- Fig. \ref{p3d}.}

In Fig. \ref{plot:contour2}, we explore the scenario when \(g_{\nu_\mu} \neq g_\mu\), and derive the upper limits on \(g_{\nu_\mu}\), while keeping all other parameters consistent with those in Fig. \ref{plot:contour}. In these plots, we consider $g_\mu=0.1$. In this scenario, the upper bounds on the coupling \(g_{\nu_\mu}\) become stringent. We find the upper bound on the self-interacting coupling parameter to be {$8.0\times10^{-6}$ and $1.4\times10^{-6}$} using the forecasted PIXIE upper bound on $\mu$ and $y$ types of spectral distortion, respectively; while using the FIRAS bound on $\mu$ and $y$ types of distortion, we get the upper limit on the coupling parameter to be {$5.7\times10^{-4}$ and $5.3\times10^{-5}$}, respectively--- when \(E_\nu=1~{\rm PeV}\)--- figure \ref{p4a}. In the right panel (Fig. \ref{p4b}), we report the bounds taking the neutrino energy to be $E_\nu= 100\ \rm{PeV}$. For this scenario, the upper bounds on the coupling parameter are {$7.5\times10^{-5}$ and $1.3\times10^{-5}$} using the forecasted PIXIE constraint on $\mu$ and $y$ types of CMB distortion, respectively. On the other hand, the upper bound becomes {$2.8\times10^{-3}$ and $5.7\times10^{-4}$} considering the FIRAS limit on $\mu$ and $y$ type distortions, respectively.

{To compare our result, we have included the bounds on coupling parameter $g_{\nu_\mu}$ by BBN, CMB anisotropy, Hubble tension, other cosmological and laboratory constraints  \cite{Huang:2017egl, PhysRevLett.123.191102, Barenboim:2019tux, Lancaster_2017, PhysRevLett.124.041802}. The sky-blue-shaded region denotes the parameter space excluded by the Big Bang Nucleosynthesis bounds on the $\Delta N_{\rm eff}$ \cite{Huang:2017egl}, where $\Delta N_{\rm eff}$ represents the extra radiation contribution to the effective number of neutrino species. The yellow- and magenta-shaded regions depict the exclusion from CMB anisotropy measurements \cite{Barenboim:2019tux}. The authors define a prior range of $\log_{10}(G_{\rm eff}\,[{\rm MeV^{-2}}])$ to $[-4.5; -0.1]$ based on the Gelman-Rubin convergence criterion. In the range $[-4.5; -2.5]$, neutrinos  behave almost free-streaming, and authors named it $\Lambda$CDM mode (magenta-shaded region in Figs. \ref{plot:contour}, \ref{plot:contour2}, and \ref{plot:contour3}). The range $[-2.5; -0.1]$, is defined as a self-interacting neutrino mode (SI mode, yellow-shaded region in Figs. \ref{plot:contour}, \ref{plot:contour2}, and \ref{plot:contour3}).} The gray-shaded region represents the excluded parameter space from the bound on the decay channel of the kaon, i.e., $K\rightarrow \mu \nu\phi$, where the kaon decays into a muon, a neutrino, and a light scalar $\phi$ \cite{PhysRevLett.124.041802}. The cyan-shaded region can explain the Hubble by considering a moderately interacting neutrino ($\mathrm{MI}\nu$) \cite{Lancaster_2017, PhysRevD.109.063007}. The green-shaded region corresponds to the parameter space where the model successfully explains the excess in muon \((g-2)\) \cite{PhysRevD.105.L051702}. Here, the value on \(g_\mu\) (\(g_{\nu_\mu}=g_\mu\) in figure \ref{plot:contour}) is obtained from the bound on effective photon-photon coupling as given in \cite{PhysRevD.105.L051702}. Moreover, this limit is valid only for \(m_\phi\ll m_\mu\), where \(m_\mu\) is the mass of the muon. In Fig. \ref{plot:contour3}, we consider the value of $g_\mu=5\times10^{-4}$ \cite{PhysRevD.105.L051702}, which explains the excess in the anomalous magnetic moment of the muon \cite{PhysRevLett.126.141801}. In this case, we get the upper bound on the coupling parameter to be {$1.5\times10^{-3}$ and $2.8\times10^{-4}$} from the PIXIE forecasted upper bounds on $\mu$ and $y$ types of CMB spectral distortion, respectively; while from the FIRAS upper limit on $\mu$ and $y$ types of distortion, we find upper bound on the coupling parameter to be {$9.0\times10^{-2}$ and $1.1\times10^{-2}$}, respectively--- for UHE neutrino energy to be $E_\nu=1$ PeV--- in figure \ref{p5a}. For UHE neutrino energy to be $E_\nu=100$ PeV, we find the upper bound on the coupling parameter to be {$1.5\times10^{-2}$ and $2.7\times10^{-3}$} using the forecasted PIXIE upper limit on $\mu$ and $y$ types of CMB distortion, respectively. While we get the upper bound on the coupling parameter to be {$3.5\times10^{-1}$ and $10^{-1}$} using the FIRAS upper limit on $\mu$ and $y$ types of distortion, respectively, as shown in the figure \ref{p5b}.

 \section{Conclusion}\label{Conclusion}
This work investigates the cosmological implications of self-interacting ultrahigh-energy (UHE) neutrinos mediated by light scalar bosons, focusing on their imprint on Cosmic Microwave Background (CMB) spectral distortions. By considering a minimal setup of the UHE neutrinos originating from decaying superheavy dark matter and interacting through a scalar, we analyzed the energy injection into the {plasma} resulting from radiative scattering between these neutrinos and the cosmic neutrino background (C$\nu$B). This process generates high-energy photons via loop-mediated interactions, which distort the CMB spectrum--- specifically producing $\mu$-type distortions at redshifts \(5 \times 10^4 \lesssim z \lesssim 2 \times 10^6\) and $y$-type distortions at \(z \lesssim 5 \times 10^4\).

Using observational constraints from COBE/FIRAS and projected sensitivities from PIXIE, we derived stringent upper bounds on the flavor-specific self-interaction coupling strength \(g_{\nu_{\mu}}\) for muon neutrinos as a function of mediator mass \(m_{\phi}\). Our key results for $f_\chi=1$ are:  
\begin{enumerate}
    \item{ For UHE neutrinos at 1 PeV, the PIXIE-projected bounds yield \(g_{\nu_{\mu}} \lesssim 1.0 \times 10^{-3}\) ($\mu$-distortion) and \(g_{\nu_{\mu}} \lesssim 3.8 \times 10^{-4}\) ($y$-distortion), improving upon FIRAS constraints by an order of magnitude.} 
    \item The bounds remain constant for \(m_{\phi} \ll \sqrt{s}\) (where \({s} \approx {2 E_{\nu_{\mu}} E_{\nu}}\) is the center-of-mass energy) but exhibit a characteristic kink near \(m_{\phi} \sim \sqrt{s}\), where the cross-section peaks. For \(m_{\phi} \gg \sqrt{s}\), the constraints relax proportionally to \(m_{\phi}\). 
    \item Increasing the UHE neutrino energy (e.g., to 100 PeV) shifts the kink to higher \(m_{\phi}\) and relaxes the bounds. 
    \item We also present a scenario with unequal couplings \(g_{\nu_{\mu}} \neq g_{\mu}\), where $g_\mu=5\times10^{-4}$ \cite{PhysRevD.105.L051702}, which explains the excess in the anomalous magnetic moment of the muon \cite{PhysRevLett.126.141801}. {In this case, the upper bounds on the coupling parameter are found to be $1.5\times10^{-3}$ ($\mu$-type) and $2.8\times10^{-4}$ ($y$-type) using the PIXIE-projected bounds, which is an improvement on FIRAS bounds by two orders of magnitude.} 
    \item {Reducing the value of \(f_\chi\) by order of \(10^1\) relaxes the constraints by a factor of approximately 1.8 when \(g_{\nu_{\mu}} = g_{\mu}\). When \(g_{\nu_{\mu}} \neq g_{\mu}\), the bounds are relaxed by a factor of about three.}
\end{enumerate}
For comparison, we have also presented existing bounds from { the BBN bounds on $\Delta N_{\rm eff}$, CMB anisotropy bounds}, rare kaon decays (\(K \to \mu\nu\phi\)), resolutions to the Hubble tension and muon \((g-2)\) anomaly. Our analysis demonstrates that CMB spectral distortions serve as a powerful, complementary probe of neutrino self-interactions beyond the Standard Model. Future missions like PIXIE could improve sensitivity by three orders of magnitude, offering unprecedented insights into neutrino physics, dark matter properties, and cosmological energy injection mechanisms.

\section {ACKNOWLEDGEMENTS} 
This work is in part supported by the Hangzhou City Scientific Research Funding, No. E5BH2B0105/B015F40725006. { We thank the referee(s) for suggestions and detailed reports that significantly improved the quality of the manuscript. A.C.N. would like to acknowledge the funding received from ISRO RAC-S, under proposal number RAC-S/GU/2024/4/56. TS is supported by the Department of Science and Technology (DST), Government of India, through the DST INSPIRE Faculty Fellowship (DST/INSPIRE/04/2024/004616). TS also received support from the National Natural Science Foundation of China under Grant Numbers 12475094, 12135006, and 12075097.}

\end{document}